\documentclass[11pt,a4paper]{article}

\usepackage{jstyle}
\usepackage{tikz}
\usepackage{amsthm,mathrsfs,stmaryrd,bm}
\usepackage[vcentermath]{youngtab}
\usepackage{simplewick}
\usepackage{framed,pifont}

\usepackage{graphics}
\usepackage{graphicx}
\usepackage{epic}
\usepackage{eepic}

\usepackage{feynmp}
\usepackage[retainorgcmds]{IEEEtrantools}
\usepackage{tensor}

\usepackage{latexsym}
\usepackage{amsmath}
\usepackage{wasysym}
\usepackage{amsfonts}
\usepackage{dsfont}
\usepackage{epsf}
\usepackage{epsfig}
\usepackage{amssymb,bm,epsf,epsfig}

\usepackage{color}
\definecolor{rougef}{rgb}{0.56,0,0}
\definecolor{vertf}{rgb}{0,0.5,0}
\definecolor{bleuf}{rgb}{0,0,0.8}
\definecolor{violetf}{rgb}{0.5,0,0.5}

\usepackage[vcentermath]{youngtab}

\renewcommand{\theequation}{\thesection.\arabic{equation}} \csname
@addtoreset\endcsname{equation}{section}

\def\Ddot{D\!\cdot\!}

\def\pe{\prime}
\def\3s{{s \choose 3}}
\def\4s{{s \choose 4}}
\def\5s{{s \choose 5}}
\def\6s{{s \choose 6}}

\def\12{\dfrac{1}{2}}

\def\nn{\nonumber}

\def\2{\ell_2}

\def\pr{\partial}

\def\prd{\partial \cdot}

\def\scri{\mathscr{I}}
\def\scrim{\mathscr{I}^-}
\def\scrip{\mathscr{I}^+}

\def\be{\begin{equation}}
\def\ee{\end{equation}}
\def\bea{\begin{eqnarray}}
\def\eea{\end{eqnarray}}
\def\ba{\begin{array}}
\def\ea{\end{array}}
\def\bec{\begin{center}}
\def\ec{\end{center}}

\def\a{\alpha} 
\def\b{\beta}  
\def\g{\gamma} 

\def\d{\delta} 

\def\e{\epsilon}

\def\h{\eta}

\def\l{\lambda}

\def\m{\mu}
\def\n{\nu}

\def\r{\rho}
\def\s{\sigma}

\def\vf{\varphi}
\def\O{\Omega}
\def\o{\omega}

\def\cA{{\cal A}}
\def\cB{{\cal B}}

\def\cF{{\cal F}}

\def\cK{{\cal K}}

\def\cO{{\cal O}}

\def\cR{{\cal R}}

\def\cT{{\cal T}}
\def\cU{{\cal U}}
\def\cV{{\cal V}}
\def\cW{{\cal W}}

\author[a]{Andrea Campoleoni,}
\author[b]{Dario Francia}
\author[b]{and Carlo Heissenberg}

\affiliation[a]{Universit{\'e} Libre de Bruxelles
and International Solvay Institutes,\\
ULB-Campus Plaine CP231,
B-1050 Brussels, Belgium} 
\affiliation[b]{Scuola Normale Superiore and INFN,\\ Piazza dei Cavalieri 7, I-56126 Pisa, Italy} 
\emailAdd{andrea.campoleoni@ulb.ac.be}  \emailAdd{dario.francia@sns.it}
\emailAdd{carlo.heissenberg@sns.it}

\title{\centering
\Huge{On higher-spin supertranslations \\
and superrotations}}

\abstract{We study the large gauge transformations of massless higher-spin fields in four-dimensional Minkowski space. Upon imposing suitable fall-off conditions, providing higher-spin counterparts of the Bondi gauge, we observe the existence of an infinite-dimensional asymptotic symmetry algebra. The corresponding Ward identities can be held responsible for Weinberg's factorisation theorem for amplitudes involving soft particles of spin greater than two.}


\begin{document}

\begin{fmffile}{diagram}

\maketitle

\setcounter{tocdepth}{1}
\tableofcontents

\section{Introduction and outlook} \label{sec: intro}

In this work we explore the possible relation between large higher-spin gauge transformations and soft theorems. Our main observation is that, upon imposing a reasonable choice of fall-off conditions at null infinity, the residual gauge transformations of Fronsdal fields \cite{fronsdal} on a four-dimensional Minkowski background generate an infinite-dimensional Abelian symmetry algebra. The Ward identities of the latter, in their turn, can be shown to reproduce the factorisation formulae of Weinberg \cite{Weinberg_64, Weinberg_65}, in strict analogy with recent results concerning the asymptotic symmetry groups of spin one and spin two gauge fields \cite{Strominger_Invariance, Strominger_Weinberg}. 

 Indeed, significant interest has been recently shown in the relation between two fairly old topics: asymptotic gravitational symmetries, discovered in the sixties by Bondi, Metzner and Sachs \cite{BMS, Sachs_Waves, Sachs_Symmetries} and later reconsidered from various perspectives (see {\it e.g.}\ \cite{Ashtekar_Asympt, Ashtekar_Symplectic, Ashtekar_Lectures, Geroch_Lectures, Barnich_BMS/CFT, Barnich_Revisited, Barnich_Charge}), and soft theorems, {\it i.e.}\ relations among scattering amplitudes for processes involving the emission or the absorption of low-energy particles \cite{Weinberg_64, Weinberg_65}. The explorations of the electrodynamical counterparts of this relation, involving the interplay between soft photons and large, spin-one gauge symmetries, pointed to the existence of a general underlying field-theoretical mechanism, thus adding further appeal to the subject. The main observation fostering the related ongoing activity is that Weinberg's soft photon and soft graviton theorems can be recast as Ward identities for suitably identified large gauge symmetries of electromagnetism and gravity, respectively \cite{Strominger_YM, Strominger_QED, Campiglia, Avery_Schwab}. 
 
Weinberg's factorisation theorem, on the other hand, holds for soft massless particles of any spin, thus naturally leading to wonder which kind of asymptotic symmetry of the corresponding gauge theories, if any, may be held responsible for it. Our purpose is to try to elucidate this point, stressing the existence of an infinite-dimensional higher-spin symmetry algebra whose Ward identities can indeed be shown to reproduce Weinberg's result. 

In Section~\ref{sec: BMSreview} we review and somehow rephrase the arguments concerning the structure of the asymptotic symmetry group for  spin-two gauge fields, exploiting for our analysis solely the structure of the linearised theory. This allows us not only to introduce our notation and general line of approach, but also to propose a derivation of Weinberg's soft theorem where the equivalence principle has not to be assumed from the very beginning (rather, it is deduced), which is relevant in view of the extension to higher spins. In Section~\ref{sec: hspst} we consider a first class of large gauge symmetries of the Fronsdal action to be identified with properly defined higher-spin supertranslations. We determine the corresponding infinite-dimensional symmetry to then show how the associated Ward identities allow to derive Weinberg's soft theorem for arbitrary integer spin. Our construction is based on the definition of a suitable Bondi-like gauge for higher spins, whose consistency is further discussed in Section~\ref{sec: Bondi}. In Section~\ref{sec: FP} we take a different perspective and consider the possibility to derive Weinberg's result for any spin as the Goldstone theorem of a specific class of large gauge transformations, thus extending the results of \cite{Ferrari_Picasso_1, Ferrari_Picasso_2}. This approach provides a non-perturbative result that allows in principle to keep track also of the subleading corrections.

Higher-spin supertranslations are actually only a particular class of the transformations preserving our Bondi-like falloff conditions. We investigate the general form of the solution in Section~\ref{sec: rot} (with some technicalities detailed in the appendices) with focus on the spin-three case, showing the existence of additional infinite families of asymptotic symmetries, providing proper higher-spin generalisations of superrotations \cite{Barnich_Revisited,Barnich_BMS/CFT,superrotations_pair-production}. The full structure of the asymptotic symmetry algebra for any value of the spin,  the computation of the corresponding charges, together with a deeper assessment of its possible role and meaning, in particular in relation with the structure of  subleading terms in soft theorems, will be explored in future work. 

In our work we uncover a new class of infinite-dimensional symmetries. This is expected to improve our insight into higher-spin theories as knowledge of additional symmetries usually does. Weinberg's soft theorems, among other consequences, imply triviality of the amplitudes involving soft higher spin quanta. (See also \cite{WWP} for a more recent analysis.) Still, we believe it to be hard to close the case concerning the relevance of massless higher spins in flat space until a clear connection with string amplitudes is established. The latter concern massive states, but it is expected that one could understand them as resulting from some massless phase with enhanced symmetry, arguably to arise after a properly defined tensionless string limit. (See e.g.\ \cite{Sagnotti_review} and references therein.) While this long-standing conjecture is usually addressed for symmetries that act everywhere in the space-time bulk, our work is meant as a first step in trying to extend the analysis at the boundary, which to our knowledge was not considered before in $D > 3$. Our ultimate hope is that these investigations may help to shed some light on the still largely mysterious infrared physics of higher-spin massless quanta.\\

\section{Soft gravitons and BMS symmetry} \label{sec: BMSreview}

In \cite{Weinberg_64, Weinberg_65}, Weinberg showed that, using only the Lorentz invariance and the pole structure of the $S$ matrix, it is possible to derive the conservation of charge and the equality of gravitational and inertial mass as consequences of the soft emission of the corresponding massless spin-one and spin-two quanta. On the same grounds, he argued that there can be no room for macroscopic  fields corresponding to particles of spin three or higher. In short, Weinberg considered  the $S$-matrix element $S_{\beta\alpha} (\mathbf q)$, for arbitrary asymptotic particle states  $\alpha \rightarrow \beta$, also involving an extra soft massless particle of $4-$momentum $q^{\, \mu}\equiv (\omega, \mathbf q) \to 0$ and helicity $s$. The two main contributions to this process are schematically encoded in  the following picture:
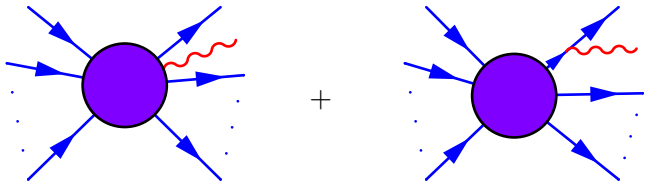
\begin{align*}
\begin{gathered}
\begin{fmfgraph*}(90, 65)
\fmfleft{i1,d1,i2,i3}
\fmfright{o1,d2,d3,o2,o4,o3}
\fmf{fermion, fore=blue}{i1,v1}
\fmf{fermion, fore=blue}{i2,v1}
\fmf{fermion, fore=blue}{i3,v1}
\fmf{fermion, fore=blue}{v1,o1}
\fmf{fermion, fore=blue}{v1,o2}
\fmf{fermion, fore=blue}{v1,o3}
\fmf{dots, fore=blue}{i2,i1}
\fmf{dots, fore=blue}{o1,o2}
\fmfset{dot_len}{4.7mm}
\fmf{photon, fore=red, tension=0}{v1,o4}
\fmfv{d.sh=circle,d.f=empty,d.si=.35w,b=(.5,,0,,1)}{v1}
\end{fmfgraph*}
\end{gathered}
\qquad + \qquad 
\begin{gathered}
\begin{fmfgraph*}(90, 65)
\fmfleft{i1,d1,i2,i3}
\fmfright{o1,d2,o2,op,o3}
\fmf{fermion, fore=blue}{i1,v1}
\fmf{fermion, fore=blue}{i2,v1}
\fmf{fermion, fore=blue}{i3,v1}
\fmf{fermion, fore=blue}{v1,o1}
\fmf{fermion, fore=blue}{v1,o2}
\fmf{fermion, fore=blue}{v1,o}
\fmf{fermion, fore=blue}{o,o3}
\fmf{photon, fore=red, tension=0}{o,op}
\fmfv{d.sh=circle,d.f=empty,d.si=.35w,b=(.5,,0,,1)}{v1}
\fmf{dots, fore=blue}{i2,i1}
\fmf{dots, fore=blue}{o1,o2}
\fmfset{dot_len}{3.95mm}
\end{fmfgraph*}
\end{gathered}
\end{align*}
The second one, in particular, provides the leading contribution to the process and takes a factorised form that, in the notation of \cite{Weinberg_64, Weinberg_65}, can be written
\be \label{wfact}
\lim_{\omega\to0^+}\omega\, S_{\beta \alpha}^{\pm s}(\mathbf q) = 
-\lim_{\omega\to0^+}\left[\, \omega\sum_i \eta_i^{\phantom{(s)}}\!\!\! g_i^{(s)} \frac{(p_i \cdot \varepsilon^{\pm}(\mathbf q))^s}{p_i\cdot q}\,\right]S_{\beta\alpha} \,,
\ee
with $\eta_i$ being $+1$ or $-1$ according to whether the particle $i$ is incoming or outgoing.

To our purposes it is useful to rewrite Weinberg's result in terms of the so-called retarded Bondi coordinates (see {\it e.g.}\ \cite{Strominger_Weinberg}),
\begin{equation} \label{BC}
t = u + r\, ,\qquad
x^1 + i x^2= \frac{2rz}{1+z\bar z}\, ,\qquad
x^3 = \frac{r(1-z\bar z)}{1+z\bar z}\, ,
\end{equation}
where $r=|\mathbf x|$.
Consider now a wave packet for a massless particle with spatial momentum centred around $\mathbf{q}$. At large times and large $r$, this wave packet becomes localised on the sphere at (null) infinity near the point 
\be
\mathbf q = {\omega}\,\widehat{\mathbf x} = \frac{\omega}{1+z\bar z}\,(z + \bar z, -i(z-\bar z), 1-z\bar z ) \, ,
\ee
so that the momentum of massless particles may be equivalently characterised by $q^\mu$ or $(\omega, z , \bar z)$.
The polarisation vectors can be chosen as follows \cite{Choi_Shim_Song}
\be\begin{aligned}\relax
\varepsilon^+(\mathbf q) &= \frac{1}{\sqrt 2}\left( \bar z, 1, -i, -\bar z\right) ,\\
\varepsilon^-(\mathbf q) &= \frac{1}{\sqrt 2}\left(  z, 1, i, - z\right) = \overline{\varepsilon^+(\mathbf q) } \, ,
\end{aligned}
\ee
thus allowing to rewrite Weinberg's soft theorem from the momentum space form \eqref{wfact} to its position-space counterpart
\be\label{WeinbergPSPACE}
\lim_{\omega\to0^+}\omega\, S_{\beta\alpha}^{+s}  = (-1)^{s}\,{2^{\frac{s}{2}-1}}
(1+z\bar z)\left[\, \sum_i \eta_i^{\phantom{(s)}}\!\!\! g_i^{(s)} 
\frac{(E_{i})^{s-1}(\bar z-\bar{z}_i)^{s-1}}{(z-z_i)(1+z_i\bar z_i)^{s-1}}
\,\right]S_{\beta\alpha} \, ,
\ee
where $E_i$ and $(z_i,\bar z_i)$ characterise the massless particles scattered to null infinity.

 For the case of spin $2$, Weinberg's soft  theorem has been recast as the Ward identity following from BMS supertranslation symmetry \cite{Strominger_Weinberg}. Moreover, it has been conjectured that infinitesimal BMS transformations provide a symmetry of both the classical gravitational scattering \emph{and} the quantum-gravitational $S$ matrix  \cite{Strominger_Invariance}. In particular, the Ward identity corresponding to this symmetry has been recognised to be the $z$-divergence of Weinberg's result \eqref{WeinbergPSPACE} for $s=2$,
under the assumption that all gravitational couplings are equal ({\it i.e.}\ $g_i^{(2)} \equiv1$), that is to say, assuming that the equivalence principle holds.

 Along an alternative path, one can infer the relevant Ward identity directly from the linearised theory via the quantum Noether theorem, as in the case of QED \cite{Avery_Schwab}. Let us go through this argument as well, so as to pave the way for our subsequent generalisation to higher spins. As an additional byproduct, we shall also be able to relate Weinberg's result to the Ward identities of supertranslations without assuming (rather, deriving, in a sense) the equivalence principle. In the following we shall make use of the coordinates  \eqref{BC}.
 
 The action for a massless Fierz-Pauli field $h_{\mu\nu}$ is
\be \label{FP}
S = \frac{1}{2}\int \mathcal E^{\mu\nu}h_{\mu\nu}\,d^Dx - \int J^{\mu\nu}h_{\mu\nu}\, d^Dx \, ,
\ee
where $\mathcal E^{\mu\nu}$ is the linearised Einstein tensor
\be
\mathcal E_{\mu\nu} = \Box h_{\mu\nu} - \partial_{(\mu} \partial\cdot h_{\nu)} - \partial_\mu\partial_\nu h' + \eta_{\mu\nu} \left(\partial\cdot\partial\cdot h - \Box h'\right)  .
\ee
Here and in the following a prime denotes a trace, while indices enclosed between parentheses are assumed to be symmetrised with the minimum number of terms needed and without normalisation factors. The Noether current associated to linearised diffeomorphisms, $\d h_{\m \n} \, = \, \pr_{(\m } \e_{\n)}$, is
\be \label{BMScurrent}
j^\mu = \frac{\delta \mathcal L}{\delta h_{\alpha\beta,\mu\nu}}\, \delta h_{\alpha\beta,\nu} - \partial_\nu \frac{\delta \mathcal L}{\delta h_{\alpha\beta,\mu\nu}}\, \delta h_{\alpha\beta} + J^{\mu\nu}\epsilon_\nu \, .
\ee
By analogy with the non-linear, asymptotically flat case (see {\it e.g.}\ \cite{Strominger_Weinberg}), we consider the following form of $h_{\m \n}$  
\be\label{eq: Bondi_gauge}
h_{\mu\nu}dx^\mu dx^\nu = \frac{2\,m_B}{r}\,du^2 - 2\, U_z du dz - 2\, U_{\bar z} du d\bar z + r\, C_{zz} dz^2 + r\, C_{\bar z \bar z}d\bar z^2 \,,
\ee
that we shall refer to as the ``Bondi gauge'',\footnote{The boundary conditions considered in literature (see {\it e.g.}\ \cite{BMS_d}) often contain other non-vanishing components of the metric with a certain fall-off behaviour. Yet, these can be  eliminated by a gauge fixing that exploits the available residual ordinary ({\it i.e.}\ non large) gauge symmetry.} to then look for the residual gauge freedom that keeps it. Notice that, by construction,
$
h' =0.
$ 
If, for simplicity, we restrict ourselves to gauge parameters $\e^\mu$  which are $u$-independent and with power-like dependence on $r$ we find a family of large gauge transformations parameterised by an arbitrary function $T(z,\bar{z})$ on the celestial sphere, that we can write in two equivalent ways as follows:\footnote{Recall $\delta h_{\mu\nu} = \partial_\mu \e_\nu + \partial_\nu \e_\mu - 2\, \Gamma^{\rho}_{\mu\nu}\e_\rho$, where the Christoffel symbols for Minkowski space in Bondi coordinates are
$$
\Gamma^z_{rz} = \frac{1}{r}\,, \qquad 
\Gamma^{z}_{zz} = \partial_z \log \gamma_{z\bar z}\,,\qquad 
\Gamma^u_{z\bar z} = r\,\gamma_{z\bar z}\,,\qquad
\Gamma^r_{z\bar z} = -\,r\,\gamma_{z\bar z}\,,
$$
while  $\g_{z\bar{z}}$ is the metric on the two-dimensional unit sphere.}
\be \label{gravity_transl}
\begin{aligned} 
\e_\mu dx^\mu &= - \left(T+ D^z D_z T \right)du - T dr - r \left( D_z T\, dz + D_{\bar z}T\, d\bar z \right) ,\\[2pt]
\e^\mu \partial_\mu  &= T \partial_u  + D^z D_z T\, \partial_r - \frac{1}{r} \left(D^zT\, \partial_z + D^{\bar z}T\, \partial_{\bar z} \right) ,
\end{aligned}
\ee
where $D_z$ is the covariant derivative on the unit 2-dimensional  sphere. In particular, the non-vanishing gauge variations are 
\begin{align}
\delta h_{uz} & = -\, D_z\!\left(\,T + D^z D_z T\,\right),\\
\delta h_{zz} &  = -\, 2\,r D^2_zT \, ,
\end{align} 
which represent infinitesimal BMS supertranslations. In this section we shall focus on these asymptotic symmetries. On the other hand, by allowing for the most general form of the residual gauge parameters $\e(u,r, z, \bar z)$ one recovers the full BMS algebra (see {\it e.g.}\ \cite{Barnich_BMS/CFT}):
\begin{equation}
\begin{aligned}
\e &= \left(T + \frac{u}{2}\, D\cdot Y\right)\partial_u
+ \left(D_zD^z T - \frac{1}{2}\left(u+r\right)D\cdot Y\right)\partial_r\\
&+ \left(Y^z - \frac{1}{r}\,D^z Y-\frac{u}{2r}\,D^zD\cdot Y\right)\partial_z
+
\left(Y^{\bar z} - \frac{1}{r}\,D^{\bar z} Y-\frac{u}{2r}\,D^{\bar z}D\cdot Y\right)\partial_{\bar z} \, .
\end{aligned}
\end{equation}
Indeed, the corresponding vector at $\scrip$ spans an infinite-dimensional family of direction-dependent translations parametrised by $T(z, \bar z)$, together with the the transformations generated by the conformal Killing vectors on the sphere $Y^z(z)$ and $Y^{\bar{z}}(\bar{z})$.

From \eqref{BMScurrent} we may now compute the leading contribution to the charge associated with the residual supertranslation gauge symmetry,
\be
Q^+ = - \int_{\scrip} T(z, \bar z)\left[\, \partial_u\! \left( D^z D^zC_{zz} + D^{\bar z} D^{\bar z} C_{\bar z \bar z} \right) + J(u, z, \bar z) \,\right] \gamma_{z\bar z}\, d^2z du \,,
\ee
where
\be
J(u, z, \bar z) \equiv \lim_{r\to\infty}r^2 J^{rr}(u, z, \bar z) \,.
\ee
We assume that supertranslations act on matter fields by $\delta\Phi(x)=i T(z, \bar z) \partial_u\Phi(x)$ at $\scrip$ and
that this action is canonically realised by $\delta \Phi(x) = i[Q, \Phi(x)]$, as shown in \cite{Strominger_Invariance}. Analogous considerations apply to $\scrim$. The correlation functions therefore satisfy
\be\begin{aligned}
	\langle \delta \prod_{n} \Phi_n(x_n) \rangle &= i\, \langle 0 | \left( Q^+ \prod_{n} \Phi_n(x_n) - \prod_{n} \Phi_n(x_n) Q^-\right) |0\rangle\\
	&= i\, \sum_{n} f_n\, T(z_n, \bar z_n) \langle\, \prod_{n} \partial_u\Phi_n(x_n)\,\rangle \,.
\end{aligned}\ee
Performing the LSZ reduction of the previous formula yields the following Ward identity (for more details see \cite{Avery_Schwab}): 
\be\label{QSSQ}
\langle\text{out}| \left(Q^+ S - S Q^-\right) |\text{in}\rangle = \sum_{i} \eta_{i}\, f_i\, E_i\, T (z_i, \bar z_i) \langle\text{out}| S |\text{in}\rangle \,,
\ee
where $Q^-$ denotes the counterpart of $Q^+$ at $\scrim$, and where  $f_i$ depends in principle on $i$ since we are not assuming that the gravitational couplings of each matter field be ruled by the equivalence principle. In analogy with \cite{Strominger_Weinberg}, we now implement the auxiliary boundary condition
\be \label{additional_boundary_spin2}
D^zD^z C_{zz} =D^{\bar z} D^{\bar z} C_{\bar z \bar z}\, \text{ at } \scri^\pm_\mp\,.
\ee
Considering also that the matter current $J$ acts trivially on the vacuum $|0\rangle$, and hence it does not contribute to the left-hand side of \eqref{QSSQ}, we effectively obtain 
\be\label{T_s=2}
Q^+ = -\,2 \int_{\scrip} T(z, \bar z) \partial_u D^z D^zC_{zz} \gamma_{z\bar z}\, d^2z du \,.
\ee
Now, in order to drive our proof to conclusion, we propose the following choice for $T(z, \bar z)$: 
\be \label{Tz_s=2}
T(z, \bar z) = \frac{1}{w-z}\frac{1+w\bar z}{1+z\bar z}\,,
\ee
so that, using
\be
\partial_{\bar z} \left( \frac{1}{w-z}\frac{1+w\bar z}{1+z\bar z} \right) = -2\pi \delta^2(z-w) + \frac{1}{2}\,\gamma_{z\bar z} \, ,
\ee
we can rewrite
\be
Q^+ = -4\pi \int D^w C_{ww}\,du + \int D^z C_{zz} \gamma_{z\bar z}\, d^2z du \, ,
\ee
where the second term is a boundary contribution on the sphere and hence gives zero.
Plugging this result, together with its counterpart at $\scrim$, into \eqref{QSSQ}, one obtains
\be\label{Ward_s=2}
-4\pi D^z\langle\text{out}|\!\left[\left(\int\! du \partial_u C_{zz} \right) S - S \left( \int\! dv \partial_v C_{zz} \right)\right]\!|\text{in}\rangle =
\sum_{i}\eta_{i}  \frac{f_i\,E_i}{z-z_i}\frac{1+z\bar z_i}{1+z_i\bar z_i} \langle\text{out}| S |\text{in}\rangle\,.
\ee
Performing the $r\to\infty$ limit, so as to express $C_{zz}$ in terms of soft graviton creation and annihilation operators, one has
\be
C_{zz} = - \frac{i}{8\pi^2} \frac{2}{(1+z\bar z)^2}\int_0^{+\infty}
d\omega_{\mathbf q} \left[ a^{\text{out}}_+(\omega_{q}\hat x) e^{-i\omega_{\mathbf q}u}
- a^{\text{out}\dagger}_-(\omega_{q}\hat x) e^{i\omega_{\mathbf q}u}\right]
\ee
and
\be
\int du\, \partial_u C_{zz} = - \frac{1}{8\pi} \frac{ 2}{(1+z\bar z)^2} \lim_{\omega\to 0^+}\left[ \omega a^{\text{out}}_+(\omega \hat x) + \omega a^{\text{out}\dagger}_-(\omega \hat x)\right].
\ee
Thus, using crossing symmetry, we also have
\be
-4\pi\, \langle\text{out}|\!\left[\left(\int\! du \partial_u C_{zz} \right) S - S \left( \int\! dv \partial_v C_{zz} \right)\right]\!|\text{in}\rangle = \frac{2}{(1+z\bar z)^2} \lim_{\omega\to 0^+}\langle \text{out}|\,\omega\, a^{\text{out}}_+(\omega \hat x) |\text{in}\rangle\,,
\ee
which implies, by comparison with \eqref{Ward_s=2},
\be
\lim_{\omega\to 0^+}\langle \text{out}|\,\omega\, a^{\text{out}}_+(\omega \hat x) |\text{in}\rangle= \lim_{\omega\to0^+} (1+z\bar z)\sum_i \eta_i f_i \frac{E_i(\bar z - \bar z_i)}{(z-z_i)(1+z_i \bar z_i)}\,,
\ee
where we have used the divergence formula
\be
\gamma_{z\bar z}\,\partial_{\bar z} \frac{2}{1+z\bar z}\sum_i \eta_i f_i \frac{E_i(\bar z - \bar z_i)}{(z-z_i)(1+z_i \bar z_i)} = \sum_i \eta_i f_i \frac{E_i(1+z\bar z_i)}{(z-z_i) (1+z_i \bar z_i)}\,.
\ee
Note that we omitted the term proportional to $\partial_{\bar z}\frac{1}{z-z_i}$ since here the delta multiplies a function which vanishes when $\bar z = \bar z_i$. This shows how the supertranslation Ward identity \eqref{Ward_s=2} implies Weinberg's factorisation formula \eqref{WeinbergPSPACE}, without assuming from the beginning $f_i=\text{constant}$.\footnote{Note also that our choice \eqref{Tz_s=2} of $T(z, \bar z)$ is not restrictive, since we can always write
$$
f(z, \bar z) = \frac{1}{2\pi}\int d^2w f(w, \bar w) \partial_{\bar w} \frac{1}{w-z}\frac{1+w\bar z}{1+z \bar z}
$$
and use the linearity of the Ward identity to recover the full supertranslation symmetry from Weinberg's theorem.}

\section{Higher-spin supertranslations} \label{sec: hspst}

Owing to the fact that Weinberg's soft theorem holds for any spin, it is natural to wonder whether the corresponding  factorisations for $s \geq 3$ should be regarded as the consequence of some infinite-dimensional symmetries acting at null infinity, as for the electromagnetic and the gravitational cases. In this section, we provide an affirmative answer to this question. We first discuss the spin-three case, to then generalise our approach to the case of arbitrary spin. For more details on the ensuing construction see \cite{carlo_tesi}.

\subsection{Spin three} \label{sec: spin3}
 
 Free spin-three gauge fields can be described by the Fronsdal action \cite{fronsdal}
\be
\frac{1}{2}\int \mathcal E^{\mu\nu\rho}\varphi_{\mu\nu\rho}\, d^Dx- \int J^{\mu\nu\rho} \varphi_{\mu\nu\rho}\,d^D\!x \,,
\ee
with the ``Einstein'' tensor $\mathcal E_{\mu\nu\rho}$ given by
\be
\mathcal E_{\mu\nu\rho} =  \mathcal F_{\mu\nu\rho}  -  \frac{1}{2}\, \eta_{ (\mu\nu}\, \mathcal F^{\, \pe}{}_{\!\rho)} \, ,
\ee
where $\cF$ is the Fronsdal, Ricci-like tensor:
\be
\mathcal F_{\mu\nu\rho} = \Box \, \varphi_{ \mu\nu\rho} - \partial_{(\mu} \partial\cdot\varphi_{ \nu\rho)} + \partial_{(\mu} \partial_{\nu\phantom{)}\!} \varphi^{\, \pe}{}_{\!\rho)} \,.
\ee
The action is invariant under the gauge symmetry
\be \label{gauge3}
\varphi_{\mu\nu\rho} \sim \varphi_{\mu\nu\rho} + \partial_{(\mu}\e_{\nu\rho)}\, ,
\ee
with the gauge parameter constrained to be traceless: $\e^{\, \pe} =0$.

By analogy with the spin-two case, we choose our  ``Bondi-like gauge'' near $\scrip$ so that the following components are assumed to vanish
\begin{alignat}{3}
& \varphi_{\, r\alpha\beta} \, = \, 0\, , \qquad & & \mbox{for all} \ \  \alpha, \, \beta; \label{bondilike1} \\
& \varphi_{\, z\bar z\m} \, = \, 0\, , \qquad & &  \mbox{for all} \ \  \m; \label{bondilike2}
\end{alignat}
while the other components scale in the following manner as $r\to\infty$ 
\be \label{boundary-4d}
\varphi_{uuu} = \frac{B}{r} \, , \qquad \varphi_{uuz} = U_z \, ,\qquad \varphi_{uzz} = r\, C_{zz} \, , \qquad \varphi_{zzz} = r^2 B_{zzz} \, ,
\ee
where $B$, $U_z, C_{zz}$ and $B_{zzz}$ are all independent of $r$, and we omitted subleading terms in $r$ in \eqref{boundary-4d}.
Analogous conditions hold for the $\bar{z}-$components. Notice that by construction $\varphi^{\, \pe}{}_\mu = 0$. Let us stress that eqs. \eqref{bondilike1}, \eqref{bondilike2} and \eqref{boundary-4d} provide a combination of gauge-fixing and scaling behaviour at $\scrip$. Equivalently, one could set to zero only fewer components of the field using the off-shell gauge symmetry and fix suitable fall-off conditions on the others. The classification of asymptotic symmetries would then hold up to subleading undetermined contributions to the gauge parameter, corresponding to ordinary residual gauge symmetries. The consistency of our boundary conditions will be further discussed in Section~\ref{sec: Bondi}.

Again, we ask ourselves whether there are residual gauge transformations, besides global Killing symmetries, leaving this structure invariant. The answer to this question is that there is indeed a residual gauge freedom given by the following family of tensors, parameterised by the arbitrary function $T(z, \bar z)$:
\be\label{supertransl_spin3}
\begin{aligned}
&\e_{\mu\nu}dx^\mu dx^\nu = 
- \left( \frac{3}{4}\,T + D^z D_z T + \frac{1}{4}\,(D^zD_z)^2T \right)du^2 
- 2 \left( \frac{3}{4}\, T + \frac{1}{4}\,D^zD_z T \right) du dr\\ 
&- 2r \left( \frac{3}{4}\,D_zT + \frac{1}{4}\,D_z^2 D^z T \right) du dz - 2r \left( \frac{3}{4}\,D_{\bar z}T + \frac{1}{4}\,D_{\bar z}^2 D^{\bar z} T \right) du d\bar z - T dr^2 \\
& - r \left( D_zT dz + D_{\bar z}T d\bar z \right) dr - \frac{r^2}{2} \left( D^2_z T dz^2 
+ D^2_{\bar z}T d\bar z^2 \right) - \frac{r^2}{2}\,\gamma_{z\bar z} \left( T + D^zD_z T \right) dz d\bar z \, ,
\end{aligned}\ee
while the corresponding contravariant tensor on $\scrip$ is given by
\be
\e^{\mu\nu}\partial_\mu \partial_\nu = -\, T(z, \bar z) \partial_u^2 \, .
\ee
This residual symmetry generalises the gravitational supertranslations \eqref{gravity_transl}. In the remainder of this section we shall explore the link between higher-spin supertranslations and Weinberg's soft theorem, while postponing to Section~\ref{sec: rot} the analysis of the full set of residual gauge symmetries of the Bondi-like gauge \eqref{bondilike1}, \eqref{bondilike2} and \eqref{boundary-4d}.

The non-vanishing gauge variations generated by \eqref{supertransl_spin3} are:
\begin{subequations}
\begin{align}
\delta \varphi_{uuz} &= - D_z \left( \frac{3}{4}\,T + D^z D_z T + \frac{1}{4}\,(D^zD_z)^2T \right) , \\
\delta \varphi_{uzz} &= - \frac{r}{2}\, D^2_z\left( 3\, T + D^z D_z T \right) , \\
\delta \varphi_{zzz} &= - \frac{3}{2}\, r^2 D^3_z T \,,
\end{align}
\end{subequations}
together with their conjugates. Like for $s=2$, the only leading contribution to the Noether charge ({\it a.k.a.} surface charge) comes from $\delta\varphi_{zzz}$,  and reads
\be \label{Q+spin3}
Q^+ = \frac{3}{4}\int_{\scrip} \gamma_{z\bar z}\,\partial_u \left[(D^z)^3 B_{zzz} + \text{c.c.}\right]  T(z,\bar z) d^2z du - \frac{3}{2}\int_{\scrip} \gamma_{z\bar z}\, J(u, z, \bar z) d^2z du \,,
\ee
where again
\be
J(u, z, \bar z) \equiv \lim_{r\to\infty}r^2 J^{rrr}(u, z, \bar z) \, .
\ee
The surface charge thus computed is in agreement with that obtainable from the results of \cite{HS-charges-cov}. Under the assumption that the residual symmetry generators act on matter fields as follows, 
\be
[Q^+, \Phi] = \frac{3}{2}\, g^{\, (3)}_i T(i\partial)^2_u \Phi \,,
\ee
where $g^{\, (3)}_i$ is the coupling of the corresponding matter field, in the frequency domain we get
\be
\langle\text{out}| (Q^+ S - S Q^-) |\text{in}\rangle = \frac{3}{2}\sum_{i} \eta_{i}^{\phantom{(3)}}\!\!\! g^{(3)}_i E^2_i T (z_i, \bar z_i) \langle\text{out}| S |\text{in}\rangle \, .
\ee
In addition, in close analogy with the condition \eqref{additional_boundary_spin2} enforced in the spin-two case, we impose the auxiliary boundary condition at $\scri^\pm_\mp$ 
\be
(D^z)^3B_{zzz} = (D^{\bar z})^3 B_{\bar z\bar z \bar z} \,.
\ee
We also leave aside the $J$ term, which again acts trivially on the vacuum, thus obtaining
\be
Q^+ = \frac{3}{2} \int_{\scrip} T(z, \bar z) \partial_u (D^z)^3 B_{zzz} \gamma_{z\bar z}\, d^2z du\, .
\ee
An analogous result holds for $Q^-$. For the function $T(z, \bar z)$ we choose a slight modification of \eqref{Tz_s=2},
\be
T(z, \bar z) = \frac{1}{w-z}\left(\frac{1+w\bar z}{1+z\bar z}\right)^{\!2},
\ee
so that, after an integration by parts in $\partial_{\bar z}$, the computation of the charge involves 
\be
\partial_{\bar z} \left( \frac{1}{w-z}\left(\frac{1+w\bar z}{1+z\bar z}\right)^{\!2}\right) =
-2\pi \delta^2(z-w) + \frac{1}{2}\,\gamma_{z\bar z}\, \frac{1+w\bar z}{1+z\bar z}\, .
\ee
Therefore
\be
Q^+ = 3\pi \int du D^w D^w B_{www} - \frac{3}{4} \int D^zD^z B_{zzz} \gamma_{z\bar z}\, \frac{1+w\bar z}{1+z\bar z}\, d^2z du\, ,
\ee
where in particular the last term is a vanishing boundary contribution.
To sum up:
\be\label{Ward_s=3}\begin{aligned}
&2\pi (D^z)^2\langle\text{out}|\!\left[\left(\int\! du \partial_u B_{zzz} \right) S - S \left( \int\! dv \partial_v B_{zzz} \right)\right]\!|\text{in}\rangle \\
&=
\sum_{i} \eta_{i}\,  \frac{g^{(3)}_i E^2_i}{z-z_i}\left(\frac{1+z\bar z_i}{1+z_i\bar z_i}\right)^{\!2} \langle\text{out}| S |\text{in}\rangle \,.
\end{aligned}\ee
The usual approximation for $B_{zzz}$ gives
\be
B_{zzz} = - \frac{i}{8\pi^2} \frac{2^{3/2}}{(1+z\bar z)^3}\int_0^{+\infty}
d\omega_{\mathbf q} \left[ a^{\text{out}}_+(\omega_{q}\hat x) e^{-i\omega_{\mathbf q}u}
- a^{\text{out}\dagger}_-(\omega_{q}\hat x) e^{i\omega_{\mathbf q}u}\right]
\ee
so that
\be
\int du\, \partial_u B_{zzz} = - \frac{1}{8\pi} \frac{ 2^{3/2}}{(1+z\bar z)^3} \lim_{\omega\to 0^+}\left[ \omega a^{\text{out}}_+(\omega \hat x) + \omega a^{\text{out}\dagger}_-(\omega \hat x)\right] .
\ee
Thus, using crossing symmetry, we also have 
\be
-4\pi \langle\text{out}|\!\left[\left(\int\! du \partial_u B_{zzz} \right) S - S \left( \int\! dv \partial_v B_{zzz} \right)\right]\!|\text{in}\rangle = \frac{2^{3/2}}{(1+z\bar z)^3} \lim_{\omega\to 0^+}\langle \text{out}|\,\omega\, a^{\text{out}}_+(\omega \hat x) |\text{in}\rangle \,,
\ee
and this implies, by comparing with \eqref{Ward_s=3},
\be
\lim_{\omega\to 0^+}\langle \text{out}|\,\omega\, a^{\text{out}}_+(\omega \hat x) |\text{in}\rangle= -\lim_{\omega\to0^+} \sqrt{2} (1+z\bar z)\sum_i \eta_i g_i\, \frac{E^2_i\, (\bar z - \bar z_i)^2}{(z-z_i)(1+z_i \bar z_i)^2} \,,
\ee
since
\be
(D^z)^2 \frac{4}{(1+z\bar z)^2}\sum_i \eta_i g_i\, \frac{E^2_i(\bar z - \bar z_i)^2}{(z-z_i)(1+z_i \bar z_i)^2} = 2 \sum_i \eta_i g_i\, \frac{E^2_i(1+z\bar z_i)^2}{(z-z_i) (1+z_i \bar z_i)^2} \,.
\ee 
This shows that the Ward identity of the residual spin-three gauge symmetry implies Weinberg's factorisation formula \eqref{WeinbergPSPACE}.

\subsection{Spin $s$} \label{sec: spins}

This section is devoted to the generalisation of the previous results to arbitrary integer spin $s$. The Fronsdal action \cite{fronsdal} is invariant under the gauge transformation 
\be
\delta \varphi_{\mu_1 \ldots \mu_s} 
= \partial_{(\mu_1} \e_{\mu_2\ldots \mu_s)} 
\ee
with a traceless gauge parameter and a doubly-traceless field. Our Bondi-like gauge is summarised by the conditions
\be \label{constr_spin-s}
\varphi_{r \mu_2 \ldots \mu_s}=0=\varphi_{z \bar z \mu_3 \ldots \mu_s}
\ee
and
\be \label{boundary_spin-s}
\varphi_{uu\ldots u \underbrace{\scriptstyle zz \ldots z}_d} = r^{d-1} B_{zz\ldots z}(u,z,\bar z) 
\ee
for $d=0,\ldots,s$, together with their conjugates. These ensure in particular that the field be traceless: $\vf^{\, \pe}_{\, \m_3 \, \ldots \m_s} = 0$.
The equations defining our residual gauge freedom, which are precisely those encoding the preservation of these scaling behaviours,
are labelled by the following numbers:
\begin{itemize}
\item the number $p$ of ``$u$'' indices appearing,
\item the number $d$ of ``$z$'' indices appearing without $\bar z$ counterpart,
\item the number $c$ of pairs ``$z \bar z$'', counted ignoring their order.
\end{itemize}
For conciseness of notation, when useful, we shall also indicate by $\varphi^p_{d ,c}$ and $\e^p_{d,c}$ the field components and the gauge parameter components, respectively, labelled with this counting criteria. 
The residual gauge freedom which preserves the given falloff conditions is independent of $u$, has power-like dependence on $r$ and satisfies the trace constraint $\e^{\, \pe} =0$. It admits the following parametrisation: 
\begin{align}\label{pd0}
\e^p_{d,0}=&- \frac{r^d D^d_zT_p(z, \bar z)}{\prod_{k=1}^d (s-p-k)}\,,\\
\label{c+1c}
\e^p_{d,c+1} =& - \frac{r^2}{2}\,\gamma_{z\bar z} \left( \e^p_{d,c} - 2\, \e^{p+1}_{d,c}\right),
\end{align}
where $T_p(z, \bar z)$ for $p=0,\ldots,s-1$ is a set of angular functions satisfying
\be
\label{p+1p}
T_{p+1} =\frac{s-p}{s[s-(p+1)]}\,T_p + \frac{1}{[s-(p+1)]^2}\,D^zD_zT_p\,.
\ee
Therefore, this family of residual gauge transformations is defined recursively in terms of only one angular function $T_0(z, \bar z) \equiv T(z,\bar z)$. The non-vanishing gauge variations are, for $s = p + d$,
\be
\delta\varphi^p_{d,0} = d\, D_{z}\e^p_{d-1,0} = -  \frac{d\,r^{d-1} D^{d}_zT_p}{\prod_{k=1}^{d-1} (s-p-k)}
\ee
which respect the $r^{d-1}$ behaviour imposed on $\varphi^p_{d,0}$. In particular
the relevant contribution to the Noether current is given by
\be
\delta \varphi_{z\ldots zz} = - \frac{s\,r^{s-1}}{(s-1)!}\,D^s_z T \, .
\ee
Using the auxiliary boundary condition $(D^z)^sB_{z\ldots zz} = (D^{\bar z})^sB_{\bar z\ldots\bar z \bar z}$ and integrating by parts, the charge corresponding to our family of large gauge transformation is therefore
\be
Q^+ = (-1)^s\frac{s}{2(s-1)!}\int_{\scrip} \partial_{\bar z} T (D^z)^{s-1}\partial_u B_{z\ldots zz} d^2z du - \frac{s}{2} \int_{\scrip} \gamma_{z\bar z} J(u, z, \bar z)d^2z du \,.
\ee
Choosing
\be
T(z, \bar z) = \frac{1}{w-z}\left(\frac{1+w\bar z}{1+z\bar z}\right)^{\!s-1}
\ee
yields
\be\begin{split}
&-4\pi\, \frac{(-1)^s}{(s-1)!}(D^z)^{s-1}\langle\text{out}|\!\left[\left(\int\! du \partial_u B_{z\ldots zz}\right) S - S \left( \int\! dv \partial_v B_{z\ldots zz}\right) \right]\!|\text{in}\rangle
\\
& = \sum_{i} \eta_i\, \frac{g_i^{(s)}E_i^{s-1}}{z-z_i}\left( \frac{1+z\bar z_i}{1+z_i\bar z_i} \right)^{\!s-1} \langle \text{out} | S | \text{in}\rangle\,,
\end{split}\ee
where we have used the action 
\be
[Q^+, \Phi] = \frac{s}{2}\,g_i^{(s)}T(i\partial_u)^{s-1}\Phi
\ee
on matter fields.
The $r\to\infty$ limit approximation gives
\be\begin{split}
&-4\pi\, \langle\text{out}|\!\left[\left(\int\! du \partial_u B_{z\ldots zz}\right) S - S \left( \int\! dv \partial_v B_{z\ldots zz}\right) \right]\!|\text{in}\rangle \\
&= \frac{2^{s/2}}{(1+z\bar z)^s}\lim_{\omega\to0^+} \left[\,\omega\langle \text{out} | a^{\text{out}}_+S | \text{in}\rangle \right]
\end{split}\ee
and hence
\be
\lim_{\omega\to0^+} \left[\,\omega\langle \text{out} | a^{\text{out}}_+S | \text{in}\rangle \right] = 
(-1)^s 2^{s/2-1} (1+z\bar z) \sum_i \eta_i\, \frac{g_i^{(s)}E_n^{s-1}}{z-z_i} \left(\frac{\bar z - \bar z_i}{1+z_i \bar z_i}\right)^{\!s-1} ,
\ee
because
\be\begin{aligned}
&\frac{1}{(s-1)!}\,(D^z)^{s-1} \frac{2^{s-1}}{(1+z\bar z)^{s-1}} \sum_i \eta_i\, \frac{g_i^{(s)}E_i^{s-1}}{z-z_i} \left(\frac{\bar z - \bar z_i}{1+z_i \bar z_i}\right)^{\!s-1} \\
&=  \sum_{i} \eta_i\, \frac{g_i^{(s)}E_i^{s-1}}{z-z_i}\left( \frac{1+z\bar z_i}{1+z_i\bar z_i} \right)^{\!s-1} \langle \text{out} | S | \text{in}\rangle \,. 
\end{aligned}\ee
Thus, Weinberg's factorisation can be understood as a manifestation of an underlying spin-$s$ large gauge symmetry acting on the null boundary of Minkowski spacetime.

\section{Consistency of the Bondi gauge} \label{sec: Bondi}

The Bondi gauge \eqref{eq: Bondi_gauge} is usually obtained from the fully nonlinear general-relativistic theory of asymptotically flat spacetimes. On the other hand, the falloff conditions on  $h_{\mu\nu}$ can be seen to result from a choice of gauge in the linearised theory, together with the requirement that the field satisfies the equations of motion asymptotically, that is at leading order in an expansion in powers of $r$. Indeed, let us first impose the gauge-fixing condition 
$
h_{r\mu}=0$,  and consider the field equations
\be
R_{\mu\nu}=\Box h_{\mu\nu}-\nabla_{\!(\mu}\nabla\cdot h_{\nu)}+\nabla_{\!\mu}\nabla_{\!\nu} h'=0 \,.
\ee
In particular, the equation for the component $R_{rr}$ in this gauge reads:
\be \label{R_rr}
\frac{2h_{z\bar z}}{r^4} + 2\,\partial_r\left(\frac{h_{z\bar z}}{r^3}\right) + \,\partial^2_r\left(\frac{h_{z\bar z}}{r^2}\right)=0 \,.
\ee
Consider then the trivial solution: $h_{z\bar z}=0$. The equation $R_{uu}=0$, taking into account the previous result, reads
\be
\frac{2}{r}\, \partial_{u}h_{uu}-\frac{2}{r^2}\,\partial_u\left(D^zh_{zu}+D^{\bar z}h_{\bar z u}\right)=0\, ,
\ee
while $R_{ru}=0$ reads
\be
\frac{2h_{uu}}{r^2}+2\,\partial_r\left(\frac{h_{uu}}{r}\right)+\partial^2_rh_{uu}- \frac{2}{r}\left(D^zh_{zu}+D^{\bar z}h_{\bar z u}\right)-\partial_r\left[\frac{1}{r^2}\left(D^z h_{zu}+D^{\bar z}h_{\bar zu}\right)\right]=0 \,.
\ee
Upon expanding $h_{uu} = 2m_B r^{\alpha}+\ldots$ and $h_{uz} = -U_z r^\beta+\ldots$, these equations yield at leading order
\be\begin{aligned}
2\,\partial_u m_B\, r^{\alpha-1}+\partial_u\left(D^zU_z+D^{\bar z}U_{\bar z}\right)r^{\beta-2}&=0\, ,\\
(\alpha+1)\left[2\alpha\, m_B+D^zU_z+D^{\bar z}U_{\bar z}\right]&=0\,.
\end{aligned}\ee 
Thus we see that the only  choices avoiding unwanted constraints on the $u$ dependence of $m_B$ and $U_z$ are either $\alpha=-1$ and $\beta=0$, or $\alpha=1$ and $\beta=2$.
We choose the ``decaying mode'' $\alpha=-1$ and $\beta=0$, thus obtaining
\be\label{eq: d_um_B}
\partial_u m_B = -\frac{1}{2}\,\pr_u \left(D^zU_z+D^{\bar z}U_{\bar z}\right) .
\ee
Taking also into account the equation for the component $R_{rz}$, \emph{i.e.}
\be
-\frac{2}{r^{3}}\,D^zh_{zz}+\frac{4}{r^2}\,h_{uz}+\left(\partial_r-\frac{2}{r}\right)\!\left( \partial_r h_{uz}+\frac{2}{r}\,h_{uz}-\frac{1}{r^2}\,D^zh_{zz}\right)=0 \,,
\ee
and
substituting $h_{zz}=C_{zz}r^\delta+\ldots$ together with the other behaviours above, we have
\be
(\delta-2)D^zC_{zz}r^{\delta-3}-2\,U_z r^{-2}=0 \,,
\ee
which imposes $\delta=1$ and
\be\label{eq: D^zC_zz}
U_z = - \frac{1}{2}\,D^z C_{zz} \,.
\ee
All in all, we recovered the falloffs
\be
h_{uu} = \frac{2m_B}{r}\,, \qquad 
h_{uz} = -\,U_z \,, \qquad
h_{zz} = r\, C_{zz} \,,
\ee
together with \eqref{eq: d_um_B} and \eqref{eq: D^zC_zz}.
One can also check that the equation $R_{uz}=0$ reduces to
\be
\partial_u\left(U_z+\frac{1}{2}\,D^zC_{zz}\right)=0
\ee
at leading order, and hence is identically satisfied in view of \eqref{eq: D^zC_zz}. One can similarly check that the remaining field equations are satisfied at leading order, consistently with the number of constraints imposed by the Bianchi identities.

For the spin-three case, let us start by imposing $\varphi_{\mu\nu r}=0$, for $\mu\nu\neq z\bar z$, which can always be achieved by exploiting the traceless gauge parameter $\e_{\mu\nu}$.
The equations of motion are
\be \label{fronsdal}
\cF_{\mu\nu\rho} \, = \, \Box \varphi_{\mu\nu\rho}-\nabla_{\!(\mu}\nabla\cdot \varphi_{\nu\rho)}+\nabla_{\!(\mu}\nabla_{\!\nu\phantom{)}\!} \varphi^{\, \pe}{}_{\!\rho)}=0 \,.
\ee
Now, $\cF_{rrr}=0$ reads
\be
\frac{2}{r^4}\,\varphi_{r z \bar z} + 2\,\partial_r \left( \frac{\varphi_{rz\bar z}}{r^3}\right) + \partial_r^2 \left( \frac{\varphi_{rz\bar z}}{r^2}\right)=0 \,,
\ee
which is consistent with $\varphi_{r z \bar z}=0$. Similarly $\cF_{\mu z \bar z}=0$ is solved by $\varphi_{\mu z \bar z}=0$. 
Now, from $\cF_{uur}=0$ we have 
\be
\begin{split}
& \frac{2}{r^2}\, \varphi_{uuu} + 
2\, \partial_r \left( \frac{\varphi_{uuu}}{r} \right) +
\partial^2_r \varphi_{uuu}-
\frac{2}{r^2}\left(D^z \varphi_{zuu} + D^{\bar z}\varphi_{\bar z uu} \right)\\
& - 
\partial_r\! \left[ \frac{1}{r^2}(D^z \varphi_{zuu}+ D^{\bar z}\varphi_{\bar zuu})\right]=0 \,,
\end{split}
\ee
and expanding $\varphi_{uuu}=B\, r^{\alpha}$, $\varphi_{zuu}=U_z\, r^{\beta}$, with $\beta=\alpha+1$, at leading order we have
\be
(\alpha+1)\left[\alpha B - (D^z U_z + D^{\bar z}U_{\bar z}) \right]=0\,.
\ee
By comparison with $\cF_{uuu}=0$, which reads
\be
\frac{4}{r}\, \partial_u \varphi_{uuu}+ \partial_u \partial_r \vf_{uuu} - \frac{3}{r^2} \left(D^z \varphi_{zuu}+D^{\bar z}\varphi_{\bar z uu} \right)=0 \,,
\ee
and yields, upon expansion,
\be
(\alpha+4)\,\partial_u B - 3\, \partial_u (D^z U_z + D^{\bar z}U_{\bar z})=0 \,,
\ee
we have two possible behaviours: a ``growing mode" $\alpha=2$, $\beta =3$ and a ``decaying mode'' $\alpha=-1$, $\beta=0$. We choose the latter, obtaining $\varphi_{uuu}=B/r$ and $\varphi_{zuu}=U_z$, together with
\be\label{eq: interr_1}
\partial_u B=  \pr_u \left(D^z U_z + D^{\bar z}U_{\bar z} \right) .
\ee
From $\cF_{zur}=0$, we have
\be
-\frac{2}{r^3}\,D^z\varphi_{zzu}+\frac{4}{r^2}\,\varphi_{zuu} + \left(\partial_r - \frac{2}{r}\right)\!\left( 
\partial_r \varphi_{zuu}+\frac{2}{r}\,\varphi_{zuu}-\frac{1}{r^2}\,D^z\varphi_{zzu}\right)=0 \,.
\ee 
Which is solved by $\varphi_{zzu}= r\, C_{zz}$ and
\be\label{eq: interr_2}
U_z = \frac{1}{2}\,D^z C_{zz}\,.
\ee
Finally, from $\cF_{zzr}=0$, we have
\be
-\frac{2}{r^3}\,D^z\varphi_{zzz}+\frac{6}{r^2}\,\varphi_{zzu}-\left(\partial_r - \frac{4}{r}\right)\!\left(\frac{1}{r^2}\,D^z \varphi_{zzz}-\frac{2}{r}\,\varphi_{zzu}-\partial_r\varphi_{zzu}\right)=0 \,,
\ee
which gives $\varphi_{zzz}=B_{zzz}\,r^{2}$ and
\be\label{eq: interr_3}
C_{zz} = \frac{1}{3}\,D^zB_{zzz}\,.
\ee
This completes the consistency check for the spin-three Bondi gauge since the number of independent equations of motion in four space-time dimensions is 10.
Notice that the relations \eqref{eq: interr_1}, \eqref{eq: interr_2} and \eqref{eq: interr_3}
are indeed preserved under the action of the residual symmetry that we found in the previous section: for instance, using $[D_{\bar z},D_z, ]D_z T=\gamma_{z\bar z}D_z T$, one can verify that the variations of $C_{zz}$ and $B_{zzz}$ satisfy \eqref{eq: interr_3}.

The consistency of the Bondi-like gauge \eqref{constr_spin-s} and \eqref{boundary_spin-s} for the spin-$s$ case can be checked in a similar manner. The equations of the form 
\be
\cF_{rr\mu_{1}\ldots\mu_{s-2}} = 0 
\ee
are identically solved once we choose $\varphi_{r\mu_{1}\ldots\mu_{s-1}}=0$ and $\varphi_{z\bar z\mu_{1}\ldots\mu_{s-2}}=0$. The equations $\cF_{ru\ldots u}=0$, $\cF_{uu\ldots u}=0$ and $\cF_{ru\ldots uz}=0$ have the same form as the analogous equations of the spin-three case obtained by removing $s-3$ indices $u$ from them: the reason is that the symbols $\Gamma^\alpha_{\beta u}$ vanish identically. The equation $\cF_{ru\ldots u z\ldots z}=0$ with a given number $1<d<s$ of $z$ indices,  reads explicitly 
\be
\frac{2}{r^3}\,D^z\varphi_{u(d+1)}- \frac{2(d+1)}{r^2}\,\varphi_{ud}-\left(\partial_r- \frac{d+2}{r}\right)\!\left(\partial_r\varphi_{ud}-\frac{1}{r^2}D^z\varphi_{u(d+1)}+\frac{2}{r}\varphi_{ud}\right)=0\,,
\ee
where for brevity $\varphi_{ud}$ denotes $\varphi_{u\ldots u z \ldots z}$ with $d$ indices $z$ and $s-d$ indices $u$.
Altogether these equations impose
\be
\varphi_{ud}=B_d\, r^{d-1}\,,
\ee
where the functions $B_d$ have to satisfy
\be
B_d = \frac{1}{d+1}\, D^zB_{d+1}\,,
\ee
whereas the other equations are identically satisfied at leading order.

\section{Soft quanta and Goldstone theorem} \label{sec: FP}

In two pioneering papers by Ferrari and Picasso \cite{Ferrari_Picasso_1, Ferrari_Picasso_2}, Weinberg's soft photon theorem \cite{Weinberg_64} and its subleading corrections \cite{Low, Kazes,Burnett} were shown to follow from the Goldstone theorem applied to the breaking of a suitable class of ``large gauge symmetries'' of QED, namely those with linear gauge parameters. The photon itself was then reinterpreted as the associated Goldstone particle. In this section we generalise the strategy of  \cite{Ferrari_Picasso_1, Ferrari_Picasso_2} to all spins. In particular we detail the case of linearised gravity, since the higher-spin case obtains by the latter in a straightforward manner, as we sketch at the end of the section. For more details see \cite{carlo_tesi}.

We work in the harmonic gauge
\be
\Box h_{\mu\nu}(x) = j_{\mu\nu}(x)\,,\qquad \partial^\mu j_{\mu\nu}(x)=0\,;
\ee
here $j_{\mu\nu}(x)$ denotes the conserved stress-energy tensor of matter together with the non-linear contributions from the Einstein equations in the ADM formulation \cite{ADM}. The tensor $j_{\mu\nu}(x)$ also generates global space-time translations via the ADM energy-momentum tensor $P_\mu$.
Consider the following family of infinitesimal local \emph{large} gauge transformations, given by the linear gauge parameter $\e_\mu(x) = -\,l_{\mu\nu} x^\nu$:
\begin{align}
\alpha(l): h_{\mu\nu}(x) &\longmapsto h_{\mu\nu}(x)-2\,l_{\mu\nu}\\ 
	        \Phi(x)  &\longmapsto \Phi(x)-if\,l_{\mu\nu} x^\mu \partial^\nu\Phi(x)\,,
\end{align}
where $f$ denotes the coupling to gravity. Taking the vacuum expectation value of $\delta^{(l)}h_{\mu\nu}$, we see that
\be
\langle \delta^{(l)}h_{\mu\nu} \rangle= -2\,l_{\mu\nu} \,.
\ee
Therefore, since the transformation $\alpha(l)$ commutes with the dynamics and the vacuum expectations are not invariant under its action, it is a broken symmetry \cite{FS_SB}. 

We turn now to the discussion of the implications of this spontaneous breaking on the spectrum of the theory. It is well-known that the breaking of an internal symmetry gives rise to massless Goldstone excitations, but in the case at hand, $\alpha(l)$ does not commute with translations:  indeed, denoting by $\tau(a)$  the action of infinitesimal translations, for $a^\mu$ a constant four-vector, we  see that
\be\begin{aligned}\relax
	[\tau(a),\alpha(l)]\Phi(x) = -i f\, a^\rho l_{\rho\sigma} \partial^\sigma \Phi(x) = f\, \tau(l_{\mu\nu}a^\nu)\Phi(x)\,.
\end{aligned}\ee
We may therefore wonder whether the Goldstone theorem still holds.
The answer to this question is affirmative: from the explicit form of the current which generates the large gauge transformations (obtained by the Noether theorem, using $\Box h_{\mu\nu} = j_{\mu\nu}$ and integrating by parts) we get
\be\label{eq: Jrho}
J^{(l)}_\rho(x) = 2\,l^{\mu\nu}\partial_\rho h_{\mu\nu}(x) - l^{\mu\nu} x_\mu \Box h_{\nu\rho}\,.
\ee
One can extend the usual proof of Goldstone's theorem \cite{FS_SB, Ferrari_Picasso_2} by using the fact that the non-covariant piece of \eqref{eq: Jrho} involves the generator $j_{\mu\nu}=\Box h_{\mu\nu}$ of the (unbroken) global symmetry.

According to the Goldstone theorem, if the symmetry is broken then there are massless one-particle modes in the Fourier transform of $\langle 0 | \delta^{(l)}B |0\rangle$, where $B$ is the order parameter. More precisely
\be\label{Goldstone2}
\lim_{R\to\infty} \langle 0 | [Q_{R,\alpha}^{(l)}, B] |0\rangle= 
\lim_{R\to\infty} \langle 0 | Q_{R,\alpha}^{(l)} E_1 B - B E_1 Q_{R,\alpha}^{(l)} |0\rangle\,,
\ee
where $E_1$ denotes the projection on zero-mass one-particle states while $R$ and $\alpha$ refer to the appropriate test functions needed to give a well-defined charge.\footnote{
The regulated charge is given by
\be \nonumber
Q_{R,\alpha}^{(l)} \equiv \int f_R(\mathbf x)\alpha(x_0)J^{(l)}_0(x) d^4\!x\,,
\ee
where the test functions $f_R$ and $\alpha$ satisfy
\be \nonumber
f_R(\mathbf x) \equiv f\left(\frac{|\mathbf x|}{R}\right),\qquad 
f(x) = \begin{cases}
	1 &\text{if }x<1\\
	0 &\text{if }x>1+\epsilon
\end{cases}
\text{ and }
\int \alpha(x_0)dx_0=1,
\ee
so that the infinitesimal variation of a local operator $B$ is 
$
i\lim_{R\to\infty}[Q_{R,\alpha}^{(l)}, B] \equiv \delta^{(l)} B.
$	
	} 
Notice that the left-hand side of the previous equation is non-vanishing if and only if the symmetry is broken.

Luckily, the non-covariant piece $l^{\mu\nu} x_\mu \Box h_{\nu0}$ gives no contribution to the right-hand side of \eqref{Goldstone2} thanks to the spectral projector $E_1$, which imposes $k^2 = 0$.
Hence, we can write
\be
\lim_{R\to\infty}\langle0|[Q_{R,\alpha}^{(l)}, B]|0\rangle = 
\lim_{R\to\infty}\int d^4x  f_R(\mathbf x) \alpha(x_0) 2l^{\mu\nu} \langle 0| \dot h_{\mu\nu}(x) E_1 B - B E_1 \dot h_{\mu\nu}(x) |0\rangle \,.
\ee
Noting that the Fourier transform of the integrand on the right-hand side is, by locality, an analytic function, we can once again rewrite this identity as follows:
\be\label{MASTER2}
\lim_{R\to\infty}\langle0|[Q_{R,\alpha}^{\mu\nu}, B]|0\rangle =  (2\pi)^{3/2} \lim_{\mathbf k\to0} 
\langle \mathbf k, \mu\nu | B |0\rangle\,,
\ee
where $Q^{\mu\nu}_{R,\alpha} l_{\mu\nu}=Q^{(l)}_{R,\alpha}$ and
\be
|\mathbf k, \mu\nu \rangle = -4i \int dx_0 \alpha(x_0) E_1 \dot{h}_{\mu\nu}(\mathbf k, x_0)|0\rangle
\ee
is the one-graviton state.
Using $B=h^{\rho\sigma}(x)$ in \eqref{MASTER2} yields
\be\label{02}
(2\pi)^{3/2} \lim_{\mathbf k\to0} 
\langle \mathbf k, \mu \nu| h^{\rho\sigma}(x) |0\rangle = -\, \eta^{\mu(\rho}\eta^{\sigma)\nu}.
\ee
Using instead $B = T(\Phi(x)\bar\Phi(0))$, where $T$ denotes time ordering, allows to recover
\be
(2\pi)^{3/2} \lim_{\mathbf k\to0} 
\langle \mathbf k, \mu\nu | T(\Phi(x)\bar\Phi(0)) |0\rangle = -\frac{if}{2}\, x^{(\mu} \partial^{\nu)} \langle 0| T(\Phi(x)\bar\Phi(0))|0\rangle\,, 
\ee
which is the Ward identity
\be\label{Wardk=0,2}
S(p) \Gamma^{\mu\nu}(p,0) S(p) = -\frac{if}{2}\, \frac{\partial}{\partial p^{\rho}}\,\eta^{\rho(\mu}p^{\nu)} S(p)\,,
\ee
where $S(p)$ is the matter field propagator and $\Gamma^{\mu\nu}(p,k)$ is the graviton-matter vertex function.
Again, if one chooses $B= T( h^{\mu_1\nu_1}(x_{1}) \ldots h^{\mu_n\nu_n}(x_n))$, in \eqref{MASTER2}
\be\begin{aligned}
&(2\pi)^{3/2} \lim_{\mathbf k\to0} 
\langle \mathbf k, \mu\nu |  T( h^{\mu_1\nu_1}(x_{1}) \ldots h^{\mu_n\nu_n}(x_n)) |0\rangle\\
& =
-\sum_{i=1}^n \eta^{\mu(\mu_i}\eta^{\nu_i)\nu}\langle 0 |T( h^{\mu_1\nu_1}(x_{1}) \ldots \widehat{h^{\mu_i\nu_i}(x_i)}\ldots h^{\mu_n\nu_n}(x_n))|0\rangle
\end{aligned}\ee
where the hat indicates that the factor has been omitted. Using the previous identity \eqref{02},
one sees that the right-hand side reconstructs the disconnected part of the left hand side, leaving as a consequence
\be
\lim_{\mathbf k\to0} 
\langle \mathbf k, \mu \nu|  T( h^{\mu_1\nu_1}(x_{1}) \ldots h^{\mu_n\nu_n}(x_n)) |0\rangle_{\text{connected}}=0\,.
\ee
Soft theorems, on the other hand, can be obtained by taking insertions of $n$ graviton fields and $2m$ matter fields
$
B = T( h^{\mu_1\nu_1}(x_{1}) \ldots \Phi(y_1)\ldots \bar \Phi(z_1)\ldots),
$
since then, again reconstructing the disconnected contributions by means of \eqref{02}, one gets
\be\begin{aligned}
&(2\pi)^{3/2} \lim_{\mathbf k\to0} 
\langle \mathbf k, \mu\nu |  T( h^{\mu_1\nu_1}(x_{1}) \ldots \Phi(y_1)\ldots \bar \Phi(z_1)\ldots) |0\rangle_{\text{connected}}\\
&=
-\frac{1}{2}\sum_{j=1}^m \left(f'_j y_j{}^{(\mu} \eta^{\nu)\rho}\frac{\partial}{\partial y_{j}^{\rho}}+ f_j z_j{}^{(\mu} \eta^{\nu)\rho}\frac{\partial}{\partial z_{j}^{\rho}}\right)\langle 0 | T( h^{\mu_1\nu_1}(x_{1}) \ldots \Phi(y_1)\ldots \bar \Phi(z_1)\ldots)  |0\rangle\,.
\end{aligned}\ee
Upon Fourier-transforming, and denoting with primes the sums with respect to the $2m+n-1$ independent momenta, we obtain
\be\begin{aligned}              
&\prod_{r,s} D(q_r)S(p'_s)K^{\mu\nu}(p, p', q) S(p_s)\\
&= \frac{1}{2} \sum'_j \left(f'_j \frac{\partial}{\partial p'_{j}{}^{\rho}}\, \eta^{\rho(\mu}p'_j{}^{\nu)} + f_j\frac{\partial}{\partial p_{j}{}^{\rho}}\, \eta^{\rho(\mu}p_j{}^{\nu)}\right)\prod_{r,s} D(q_r) S(p'_s) K(p,p',q)S(p_s)\,,
\end{aligned}\ee
where $D(q)$ is the graviton propagator, and $K^{\mu\nu}(p,p',q)$ denotes the amputated amplitude for the process $K(p,p',q)$ with the addition of an extra soft graviton with momentum $k^\mu$;
taking also into account the Ward identity \eqref{Wardk=0,2} when applying the derivatives on the right-hand side gives 
\be
\begin{split}
K^{\mu\nu}(p,p',q)\label{WeinbergPoles2}
&=
i\sum_{j=1}^m \left\{ \Gamma^{\mu\nu}(p'_j, 0) S(p'_j) + S(p_j)\Gamma^{\mu\nu}(p_j,0)\right\} K(p,p',q) \\
&+\frac{1}{2}\sum'_j
\left(f'_j \frac{\partial}{\partial p'_{j}{}^{\rho}}\,\eta^{\rho(\mu} p'_{j}{}^{\nu)}+ f_j \frac{\partial}{\partial p_{j}{}^{\rho}}\, \eta^{\rho(\mu}p_j{}^{\nu)}\right)
\prod_{r,s}  K(p,p',q) \, .
\end{split}
\ee
The first line of \eqref{WeinbergPoles2} encodes Weinberg's poles as can be easily seen by considering, for instance,
\be
\begin{aligned}
\Gamma^{\mu\nu}(p', k) S(p' + k) \sim \frac{p'^\mu p'^\nu}{-(p'+k)^2+m^2} \sim \frac{p'^\mu p'^\nu}{p'\cdot k}
\end{aligned}
\ee
and is associated with those diagrams where the soft graviton is emitted or absorbed by an external line; the second line, on the other hand, encodes finite corrections  corresponding to the other diagrams, analogous to those discussed for QED and gravity in {\it e.g.}\ \cite{Low, Kazes, Burnett, soft_QED_Strominger} and \cite{soft-subleading}. 

The previous discussion can be naturally extended in the context of spin-$s$ gauge theories, by choosing the higher-spin de Donder gauge,
\be
\partial\cdot\varphi_{\mu_2\ldots\mu_s} = \frac{1}{2}\, \partial_{(\mu_2}\varphi^{\, \pe}{}_{\!\mu_3\ldots\mu_s)}\,.
\ee
Limiting ourselves to the main formulae, the Ward identity linking the matter propagator $S(p)$ to the $s$-field vertex function $\Gamma^{\mu_1\ldots\mu_s}$ reads
\be\label{Wardk=0,s}
S(p) \Gamma^{\mu_1\ldots\mu_s}(p,0) S(p) = -\frac{i}{s}\,g^{(s)} \frac{\partial}{\partial p^{\rho}}\, \eta^{\rho(\mu_1}p^{\mu_2}\ldots p^{\mu_s)} S(p)\,,
\ee
and the soft theorem expressing the amputated amplitude $K^{\mu_1\ldots\mu_s}(p,p',q)$ for the process $K(p,p',q)$ with the addition of an extra soft spin-$s$ particle with momentum $k^\mu$ is encoded in the following expression,
\be
\begin{split}
& K^{\mu_1\ldots\mu_s}(p,p',q)\label{WeinbergPoless}
=i\sum_{j=1}^m \left\{ \Gamma^{\mu_1\ldots \mu_s}(p'_j, 0) S(p'_j) + S(p_j)\Gamma^{\mu_1\ldots\mu_s}(p_j,0)\right\} K(p,p',q) \\
&+\frac{1}{s}\sum'_j 
\left( g'_j{}^{(s)}\frac{\partial}{\partial p'_{j}{}^{\rho}}\, \eta^{\rho(\mu_1} p'_j{}^{\mu_2}\ldots p'_j{}^{\mu_s)} +
g_j^{(s)}\frac{\partial}{\partial p_{j}{}^{\rho}}\, \eta^{\rho (\m_1}p_j{}^{\mu_2}\ldots p_j{}^{\mu_s)}\right)
K(p,p',q) \,.
\end{split}
\ee
Let us stress that in close parallel to the low-spin case, also for spin $s$ we obtain from the first line Weinberg's factorisation theorem for a  spin-$s$ soft particle, while the remaining terms encode subleading corrections, whose detailed analysis we postpone to future work.

\section{Higher-spin superrotations: the spin-3 example} \label{sec: rot}

In Section~\ref{sec: hspst} we identified asymptotic symmetries that suffice to recover Weinberg's factorisation theorem from the associated Ward identities. In this section we show that the full set of residual gauge transformations leaving the boundary conditions invariant is actually much larger. Indeed, the additional symmetries are generated by a number of holomorphic and antiholomorphic functions, thus generalising the local infinite-dimensional enhancement of the Lorentz algebra observed in gravity \cite{Barnich_Revisited,Barnich_BMS/CFT}. For simplicity, we illustrate this phenomenon by focussing on a field of spin three. We generalise the Bondi-like gauge of Section~\ref{sec: spin3} and the analysis of asymptotic symmetries to any number of space-time dimensions. This approach allows to better appreciate  how the infinite-dimensional enhancement appears to be a peculiarity of four-dimensional Minkowski space.

\subsection{Boundary conditions reloaded}

We parameterise the Minkowski background as follows:
\be
ds^2 = -\, du^2 - 2 dudr + r^2 \g_{ij}(x^k) dx^i dx^j \, ,
\ee
where $\g_{ij}$ denotes the metric on the unit celestial sphere of dimension $n$. The key of the Bondi-like gauge proposed in Section~\ref{sec: spin3} lies in the choices
\be \label{basic-cond}
\vf_{r\a\b} = 0 \, , \qquad
g^{\n\r} \vf_{\m\n\r} = 0 \, .
\ee  
The number of conditions that one imposes in this way is the same as in the transverse-traceless gauge, which is  reachable on shell for any value of the spin. Therefore we assume that the conditions \eqref{basic-cond} can be imposed for any $n$ on field configurations satisfying Fronsdal's equations asymptotically (as the gravity Bondi gauge does) and that possible deviations are suppressed at null infinity so as to become irrelevant for the analysis of asymptotic symmetries.\footnote{Equivalently, as discussed in section \ref{sec: spin3}, \eqref{basic-cond} can be considered as the result of a complete fixing of the ordinary gauge symmetry not affecting the physical state of the system.}

The power-like radial dependence of the remaining components ---~specified in \eqref{boundary-4d} when the dimension of space-time is equal to four, that is for $n=2$~--- is such that $\vf_{uuu}$ shares the same leading exponent as the one of the deviation $h_{uu}$ from the background metric in gravity, while the other leading exponents grow by one unity for any additional angular index on the celestial sphere. As shown in Section~\ref{sec: Bondi}, these conditions guarantee that the fields satisfy the linearised equations of motion at leading order. Following the same reasoning, one can generalise \eqref{boundary-4d} as follows: 
\begin{subequations} \label{boundary}
\begin{align}
\varphi_{uuu} & = r^{-\frac{n}{2}}\, B(u,x^l) + \cO(r^{-\frac{n}{2}-1}) \, , \\[2pt]
\varphi_{uui} & = r^{1-\frac{n}{2}}\, U_i(u,x^l) + \cO(r^{-\frac{n}{2}})\, , \\[2pt]
\varphi_{uij} & = r^{2-\frac{n}{2}}\, C_{ij}(u,x^l) + \cO(r^{1-\frac{n}{2}})\, , \\[2pt]
\varphi_{ijk} & = r^{3-\frac{n}{2}} B_{ijk}(u,x^l) + \cO(r^{2-\frac{n}{2}})\, ,
\end{align}
\end{subequations}
where, in order to satisfy \eqref{basic-cond}, the tensors $C_{ij}$ and $B_{ijk}$ are bound to be traceless,
\be \label{traceB}
\g^{ij} C_{ij} = \g^{jk} B_{ijk} = 0 \, .
\ee

The falloff of $\vf_{uuu}$ corresponds to the ``standard'' falloff of $h_{uu}$ in asymptotically flat solutions of higher-dimensional gravity \cite{BMS_d,BMS-memory}.\footnote{Alternative boundary conditions for gravity ---~designed to keep supertranslations also when the dimension of space-time is larger than four~--- have been proposed in \cite{alternative-bnd}. A similar option may be foreseen for higher spins too; we postpone an analysis of this issue to future work. Here we employ boundary conditions affine to those usually considered in literature for gravity, implementing the idea that fields should falloff faster at infinity with the increasing of the dimensionality of space-time.}  As detailed in Appendix \ref{sec:eom}, fields behaving in this way at null infinity solve Fronsdal's equations at leading order in an expansion in powers of $r$ for any value of $n$,\footnote{In complete analogy, the boundary conditions that give finite higher-spin charges in AdS also satisfy the field equations asymptotically \cite{charges-AdS,fermi-charges-AdS}. In three space-time dimensions these falloffs have also been proved to remain valid even when interactions are switched on \cite{asymptotics-metric-3d}.} provided that the following relations hold:
\be \label{rel-tens}
B = \frac{2}{n}\, D^i U_i \, , \qquad
U_i = \frac{2}{n+2}\, D^j C_{ij} \, , \qquad
C_{ij} = \frac{2}{n+4}\, D^k B_{ijk} \, ,
\ee
where $D_i$ denotes the covariant derivative on the celestial sphere. The previous discussion formally applies also to odd space-time dimensions, which the analysis of \cite{BMS_d,BMS-memory} does not encompass, while in the special $n=2$ case the first relation in \eqref{rel-tens} is substituted by the weaker condition \eqref{eq: interr_1}, involving a $u$-derivative of the same tensors.

Analogy with the gravitational falloffs and consistency with the linearised field equations are our main motivations for imposing the boundary conditions \eqref{basic-cond}--\eqref{rel-tens}, where one could also adopt a conservative viewpoint and bound the value of $n$ to be even in analogy with \cite{BMS_d,BMS-memory,alternative-bnd}.\footnote{In three space-time dimensions, asymptotic symmetries for higher-spin fields in Minkowski space have been studied in the Chern-Simons formulation \cite{3D-flat_1,3D-flat_2}, while the relation between $\scri^\pm_\mp$ has been studied in \cite{Prohazka:2017equ}. Our fall-off conditions \eqref{boundary} differ from the metric-like translation of the Chern-Simons boundary conditions displayed in \cite{3D-flat_2}. To gain a better grasp on possible subtleties emerging when the number of space-time dimensions is odd, it will be interesting to analyse how the proposal of \cite{3D-flat_2} may fit into the previous discussion.} We are now going to identify the residual gauge transformations leaving them invariant.

\subsection{Higher-spin superrotations}

The conditions $\d \vf_{r\m\n} = 0$ fix the radial dependence of all components of the traceless gauge parameter $\e^{\m\n}$. The first three conditions in
\be \label{admissible-var}
\d \vf_{uuu} = \cO(r^{-\frac{n}{2}}) \, , \quad
\d \vf_{uui} = \cO(r^{1-\frac{n}{2}}) \, , \quad
\d \vf_{uij} = \cO(r^{2-\frac{n}{2}}) \, , \quad
\d \vf_{ijk} = \cO(r^{3-\frac{n}{2}})
\ee
fix instead the dependence on $u$. All in all, the previous constraints, together with the constraint $g_{\m\n} \e^{\m\n} = 0$, are satisfied by 
\begin{align}
\e^{ij} & = \left[ K^{ij} + \frac{u}{r}\, \cT_2^{\,ij}(K) + \left(\frac{u}{r}\right)^{\!2} \cT_{4}^{\,ij}(K) \right] + \frac{1}{r} \left[\, \cU_{1}^{ij}(\r) + \frac{u}{r}^{\phantom{2}}\! \cU_{3}^{ij}(\r) \right] + \frac{1}{r^2}\, \cV_{2}^{ij}(T) \, , \label{e^ij} \\[2pt]
\e^{ui} & = \frac{u}{n+2} \left[ \Ddot K^i - \frac{u\,r^{-1}}{2(n+1)}\, D^i \Ddot \Ddot K \right] - \left[ \r^i - \frac{u\,r^{-1}}{n+1}\, D^i \Ddot \r \right] + \frac{1}{2\,r} \, D^i T \, , \label{e^ui} \\[2pt]
\e^{uu} & = \frac{u^2}{(n+1)(n+2)}\, \Ddot\Ddot K - \frac{2\,u}{n+1}\, \Ddot\r - T \, . \label{e^uu}
\end{align}
Before presenting the corresponding radial components, let us stress that the key point of the whole analysis is that the residual symmetry is parameterised by the tensors $T(x^k)$, $\r^i(x^k)$ and $K^{ij}(x^k)$ defined on the celestial sphere. They  appear at $\cO(r^0u^0)$ respectively in $\e^{uu}$, $\e^{ui}$ and $\e^{ij}$ and they must satisfy some differential constraints that will be specified below. The combinations $\cT_A^{\,ij}(K)$, $\cU_{A}^{ij}(\r)$ and $\cV_{2}^{ij}(T)$ ---~where the subscript denotes the order of the  differential operators involved~--- are instead displayed in Appendix \ref{sec:details}. The tensors $T$, $\r^i$ and $K^{ij}$ completely specify also the radial components of the gauge parameter as follows:
\begin{align}
\e^{ri} & = -\, \frac{1}{n+2} \left\{ r \cA^i(K) + \cB^i(\r) - \frac{1}{2\,r}\, D^i\! \left( \Box + n \right) T \right\} , \label{e^ri} \\[2pt]
\e^{ru} & =  -\, \frac{u\,r}{(n+1)(n+2)} \left( 1 + \frac{u}{r} \right)\! \Ddot\Ddot K + \frac{r}{n+1} \left[ \Ddot\r - \frac{u\,r^{-1}}{n+2} \left( \Box -2 \right) \right]\! \Ddot\r \nn \\
& \phantom{=}\ - \frac{1}{2(n+2)} \left( \Box - 2 \right) T \, , \label{e^ru} \\[2pt]
\e^{rr} & = \frac{r^2}{(n+1)(n+2)} \left( 1 + \frac{u}{r} \right)^{\!2}\! \Ddot\Ddot K + \frac{2\,r}{(n+1)(n+2)} \left( 1 + \frac{u}{r} \right)\! \left( \Box + n \right)\! \Ddot\r \nn \\
& \phantom{=}\ - \frac{1}{2n(n+2)} \left( \Box + n \right)\! \left( \Box + 2 \right) T \, , \label{e^rr}
\end{align}
where $\cA^i(K)$ and $\cB^i(\r)$ are given in Appendix \ref{sec:details}. 

Gauge parameters of this type induce variations of the fields such that $\d \vf_{r\m\n} = 0$ and $\d \vf_{uuu} = 0$, while 
\begin{align}
\d \vf_{uui} & = -\,\frac{1}{12n(n-1)}\, \Ddot\Ddot \cT_i \, , \label{phi_uui} \\[5pt]
\d \vf_{uij} & = \frac{r^2}{2(n+1)}\, \Ddot \cR_{ij} - \frac{u\,r}{(n+1)(n+2)} \left( D_{(i} D_{j)} + 2\,\g_{ij} \right) \left( \Box + 2(n+1) \right) \Ddot \r \nn \\
& - \frac{r}{6n}\, \Ddot \cT_{ij} \, , \label{phi_uij} \\[5pt]
\d \vf_{ijk} & = r^4\, \cK_{ijk} + \frac{u\,r^3}{n} \left\{ \Box \cK_{ijk} - D_{(i} \Ddot \cK_{jk)} + \frac{1}{n+1}\, \g_{(ij} \Ddot\Ddot \cK_{k)} + (n-3)\, \cK_{ijk} \right\} \nn \\
& + \frac{u^2 r^2}{4(n+1)(n+2)} \left( D_{(i} D_j D_{k)} + 8\, \g_{(ij} D_{k)} \right) \Ddot\Ddot K \nn + r^3\, \cR_{ijk} \\
& - \frac{u\,r^2}{2(n+1)} \left( D_{(i} D_j D_{k)} - \frac{2}{n+2}\, \g_{(ij} D_{k)}\! \left( 3\Box + 2 (n-1) \right) \right) \Ddot\r + \frac{r^2}{4}\, \cT_{ijk} \, , \label{phi_ijk}
\end{align}
where we introduced the traceless tensors
\begin{align}
\cK^{ijk} & \equiv D^{(i} K^{jk)} - \frac{2}{n+2}\, \g^{(ij} \Ddot K^{k)} \, , \label{K} \\[5pt]
\cR^{ijk} & \equiv D^{(i} D^j \r^{k)} - \frac{2}{n+2} \left(\, \g^{(ij} \Box \r^{k)} + \g^{(ij} \big\{ D^{k)} , D^l \big\} \r_l \,\right) , \label{R} \\[5pt]
\cT^{ijk} & \equiv D^{(i} D^j D^{k)} T - \frac{2}{n+2} \left(\, \g^{(ij} \Box D^{k)} T + \g^{(ij} \big\{ D^{k)} , D^l \big\} D_l T \,\right) . \label{T}
\end{align}
We recall that, as in previous sections, indices enclosed between parentheses are assumed to be symmetrised by using the minimum number of terms needed and without normalisation factor.

Consistently with the analysis of Section~\ref{sec: spin3}, when $n = 2$ the boundary conditions \eqref{basic-cond}--\eqref{rel-tens} are preserved by gauge transformations generated by an arbitrary function $T$, while $\r^i$ and $K^{ij}$ must satisfy
\be \label{constr-KR}
\cK^{ijk} = 0 \, , \qquad \cR^{ijk} = 0 \, .
\ee
When $n > 2$, these constraints still apply, while with our choice of boundary conditions $T$ also has to satisfy
\be \label{constr-T}
\cT^{ijk} = 0 \, .
\ee
The constraints \eqref{constr-KR} (plus \eqref{constr-T} when it is relevant) suffice to preserve the boundary conditions \eqref{boundary}: indeed, in any number of space-time dimensions the variations of the traces of $\vf_{uij}$ and $\vf_{ijk}$ read
\begin{align}
\g^{ij} \d \vf_{uij} & = - \frac{u\,r}{3(n-1)(n+1)}\, \Ddot\Ddot\Ddot \cR \, , \\[5pt]
\g^{jk} \d \vf_{ijk} & = - \frac{u\,r^3}{n+1} \left\{ \Ddot\Ddot \cK_i - \frac{u}{2n\,r}\, D_i \Ddot\Ddot\Ddot \cK \right\} . 
\end{align}
Moreover, the identities \eqref{identity_1}--\eqref{identity_3} guarantee the preservation of the boundary conditions when $n > 2$, by ensuring the cancellation of the terms that cannot be anymore interpreted as variations of $U_i$, $C_{ij}$ and $B_{ijk}$. Notice also that, consistently with the relations \eqref{rel-tens} induced by the requirement that the linearised field equations be satisfied at leading order, when $n=2$ one has
\begin{align} 
0 & = \pr_u D^i \d U_i \, , \label{var-rel-tens} \\[5pt]
\d U_i & = \frac{1}{2}\, D^j \d C_{ij} + \frac{u}{18}\, D_i \Ddot\Ddot\Ddot \cR \, , \\[5pt]
\d C_{ij} & = \frac{1}{3}\, D^k \d B_{ijk} - \frac{u^2}{72}\, D_{(i} D_{j)} \Ddot\Ddot\Ddot\cK \nn \\
& - \frac{1}{18} \left( \Box \Ddot \cR_{ij} - D_{(i} \Ddot\Ddot \cR_{j)} + 2\,\g_{ij} \Ddot\Ddot\Ddot \cR - 2\,\Ddot\cR_{ij} \right) \, .
\end{align}

The characterisation of asymptotic symmetries therefore reduces to the classification of the solutions of the equations \eqref{constr-KR} (plus \eqref{constr-T} when $n>2$).  First of all, let us stress that, for any value of $n$, these equations must admit a number of independent solutions that is greater than or equal to the number of traceless rank-2 Killing tensors of Minkowski space. In Cartesian coordinates, the latter indeed satisfy the equations
\be \label{killing}
\pr_{(\m} \e_{\n\r)} = 0 \, , \qquad g_{\m\n} \e^{\m\n} = 0 \, ,
\ee
which are just a particular instance of the problem at stake and are solved by
\be \label{sol-killing}
\e_{\m\n} = A_{\m\n} + A_{\m\n|\r}\, x^\r + A_{\m\n|\r\s}\, x^\r x^\s \, ,
\ee
where the involved tensors are traceless and irreducible, that is $A_{(\m\n|\r)} = A_{(\m\n|\r)\s}=0$. This implies that \eqref{killing} admit $\frac{n(n+3)(n+4)(n+5)}{12}$ independent solutions. When $n > 2$, solving either \eqref{admissible-var} or \eqref{killing} actually imposes the same conditions on the gauge parameters.
In Appendix \ref{sec:details} we also verify explicitly that the number of solutions of \eqref{constr-KR} and \eqref{constr-T} agrees with that of \eqref{killing}, at least when one considers the flat limit of the former.

On the contrary, when $n=2$ the function $T(x^k)$ is not constrained at all if one only demands preservation of the Bondi-like gauge. This leads to the higher-spin supertranslations discussed in Section~\ref{sec: spin3}. The tensors $\r^i$ and $K^{ij}$ are instead still bounded to satisfy the differential equations \eqref{constr-KR}. Remarkably, when $n = 2$, locally they both admit infinitely many solutions. This is well known for the first equation in \eqref{constr-KR}, which is the rank-2 conformal Killing equation \cite{conformal-killing}. Being traceless, it only admits two non-trivial components that, using a holomorphic parameterisation of the metric, read 
\be
\pr_{\bar{z}} K^{zz} = 0 \, , \qquad 
\pr_{z} K^{\bar{z}\bar{z}} = 0 \, .
\ee
Its solutions are therefore locally characterised by a holomorphic and an antiholomorphic functions:
\be \label{sol2-K}
K^{zz} = K(z) \, , \qquad 
K^{\bar{z}\bar{z}} = \tilde{K}(\bar{z}) \, , \qquad
K^{z\bar{z}} = 0 \, .
\ee
In a similar fashion, the second traceless equation in \eqref{constr-KR} only admits two non-trivial components that one can cast in the form
\be
\pr_{\bar{z}}\! \left( \g^{z\bar{z}} \pr_{\bar{z}} \r^z \right) = 0 \, , \qquad
\pr_{z}\! \left( \g^{z\bar{z}} \pr_{z} \r^{\bar{z}} \right) = 0 \, .
\ee
These equations are solved by
\be \label{sol2-r}
\r^z = \a(z)\, \pr_z k(z,\bar{z}) + \b(z) \, , \qquad
\r^{\bar{z}} = \tilde{\a}(\bar{z})\, \pr_{\bar{z}} k(z,\bar{z}) + \tilde{\b}(\bar{z}) \, ,
\ee
where $k(z,\bar{z})$ is the K\"ahler potential for the $2$-dimensional metric on the unit sphere. For instance, in the coordinates \eqref{BC} one has $k(z,\bar{z}) = 2\log(1+z\bar{z})$. $\a(z)$ and $\b(z)$ are instead arbitrary holomorphic functions and similar considerations apply to the antiholomorphic sector.

To conclude, we wish to sketch a possible interpretation of the infinite-dimensional families of symmetries we found. In the case of gravity supertranslations can be considered as an infinite-dimensional enhancement of the Poincar\'e translation symmetry generated, say, by $P^i$. Similarly, superrotations correspond to an infinite-dimensional enhancement of the Lorentz symmetry generated by $M^{ij}$. As discussed in Appendix \ref{sec:details}, the global solutions of the constraints \eqref{constr-KR} and \eqref{constr-T} are in one-to-one correspondence with the traceless projections of the combinations $P^{(i}P^{j)}$, $P^{(i}M^{j)k}$ and $M^{k(i}M^{j)l}$. After a proper specification of the involved representation, these products are expected to be identified with the spin-three generators of a would-be higher-spin algebra, if any, with Poincar\'e subalgebra (see {\it e.g.}\ \cite{flat-algebras_1,flat-algebras_2} for  discussions on higher-spin algebras possibly related to four-dimensional Minkowski space). The asymptotic symmetries generated by $T$, $\r^{i}$ and $K^{ij}$ can thus be interpreted as the infinite-dimensional enhancement of the Killing symmetries associated, respectively, to the products of Poincar\'e generators $P^{(i}P^{j)}$, $P^{(i}M^{j)k}$ and $M^{k(i}M^{j)l}$.  Certainly, controlling better this relation will require to consider interactions, in order to capture possible non-Abelian deformations of the asymptotic symmetry algebra. Let us anyway notice that, up to the signature, the $M^{ij}$ also generate the algebra of isometries of AdS$_3$. Their products ---~or, more precisely, a proper quotient of the universal enveloping algebra of the Lorentz algebra~--- should then give a subalgebra isomorphic to one of the AdS$_3$ higher-spin algebras that appeared in the literature. The latter are generically asymptotically enhanced to $\cW$-symmetries \cite{W1,W2,W3,W4}, which are generated by the conformal Killing tensors of the two-dimensional boundary of AdS$_3$ \cite{asymptotics-metric-3d}. The compelling similarity with the symmetry generated by $K^{ij}$ suggests that any possible non-Abelian deformation of our algebra of asymptotic symmetries should contain an infinite-dimensional non-linear $\cW$-algebra as a subalgebra.


\acknowledgments

We would like to thank G.~Barnich, H.~Godazgar, B.~Oblak and F.~Strocchi for very useful discussions and comments. This work was supported in part by Scuola Normale Superiore. The work of D.F. was also partly supported by INFN (I.S. Stefi), and  by the Munich Institute for Astro- and Particle Physics (MIAPP) of the DFG cluster of excellence ``Origin and Structure of the Universe''. A.C. acknowledges Scuola Normale Superiore and INFN Sezione di Pisa for hospitality, and his work has been partially supported by the ERC Advanced Grant ``High-Spin-Grav'' and by FNRS-Belgium (convention FRFC PDR T.1025.14 and convention IISN 4.4503.15).

\begin{appendix}

\section{Spin-3 linearised field equations in any dimension}\label{sec:eom}

When imposing the conditions \eqref{basic-cond}, the components $\cF_{rrr}$, $\cF_{rru}$ and $\cF_{rri}$ of the Fronsdal tensor \eqref{fronsdal} vanish identically. The remaining radial components take instead the form
\begin{align}
\cF_{ruu} & = \frac{1}{r^2} \left\{ \left( r^2 \pr_r^2 + n\,r\pr_r \right) \vf_{uuu} - \pr_r D^i \vf_{uui} \right\} , \label{R_ruu} \\[5pt]
\cF_{rui} & = \frac{1}{r^2} \left( r^2 \pr_r^2 + (n-2)\, r\pr_r - 2(n-1) \right) \vf_{uui} - \frac{1}{r^3} \left( r\pr_r -2 \right) D^j \vf_{uij} \, , \label{R_rui} \\[5pt]
\cF_{rij} & = \frac{1}{r^2} \left( r^2 \pr_r^2 + (n-4)\, r\pr_r - 4(n-1) \right) \vf_{uij} - \frac{1}{r^3} \left( r\pr_r - 4 \right) D^k \vf_{ijk} \, . \label{R_rij}
\end{align}
The components without radial indices also involve derivatives in the retarded Bondi time $u$. Those with at least one $u$ index read
{\allowdisplaybreaks
\begin{align}
\cF_{uuu} & = \frac{1}{r} \left\{ \left( r\pr_r + 2n \right) \pr_u\vf_{uuu} - 3\,r^{-1} \pr_u D^i \vf_{uui} \right\} + \frac{1}{r^2} \left( \Box + r^2 \pr_r^2 + n\,r\pr_r \right) \vf_{uuu} \, , \label{R_uuu} \\[5pt]
\cF_{uui} & = \frac{1}{r} \left\{ (n+2)\,\pr_u \vf_{uui} - 2\,r^{-1} \pr_u D^j \vf_{uij} + \left( r\pr_r + (n-2) \right) \pr_i \vf_{uuu} \right\} \nn \\*
& + \frac{1}{r^2} \left\{ \left( \Box + r^2 \pr_r^2 + (n-2)\,r\pr_r - (n-1) \right) \vf_{uui} - D_i D^j \vf_{uuj} \right\} , \label{R_uui} \\[5pt]
\cF_{uij} & = -\, \frac{1}{r} \left\{ \left( r\pr_r - 4 \right) \pr_u \vf_{uij} + r^{-1}\pr_u D^k \vf_{ijk} \right\} \nn \\*
& + \frac{1}{r^2}\, \Big\{\! \left( \Box + r^2\pr_r^2 + (n-4)\, r\pr_r - 2(n-2) \right) \vf_{uij} - D_{(i} D^k \vf_{j)ku} \nn \\*
& + r \left[ \left( r\pr_r + (n-2) \right) D_{(i} \vf_{j)uu} + 2\, \g_{ij} D^k \vf_{uuk} \right] - 2\,r^{2}\, \g_{ij} \left( r\pr_r + (n-1) \right) \vf_{uuu} \Big\} \, . \label{R_uij}
\end{align}
}
Finally, the component with all indices valued on the celestial sphere reads 
\be \label{R_ijk}
\begin{split}
\cF_{ijk} & = -\, \frac{1}{r} \left( 2\,r\pr_r + (n-6) \right) \pr_u \vf_{ijk} \\
& + \frac{1}{r^2}\, \Big\{\! \left( \Box + r^2\pr_r^2 + (n-6)\, r\pr_r - 3(n-3) \right) \vf_{ijk} - D_{(i} D^l \vf_{jk)l} \\
& + r \left[ \left( r\pr_r + (n-2) \right)\! D_{(i} \vf_{jk)u} + 2\, \g_{(ij} D^l \vf_{k)lu} \right]\! - 2\,r^2 \left( r\pr_r + (n-1) \right) \g_{(ij} \vf_{k)uu} \Big\} \, .
\end{split}
\ee

An ansatz of the form
\be \label{ansatz-alpha}
\vf_{uuu} = r^\a B \, , \quad
\vf_{uui} = r^{\a+1} U_i \, , \quad
\vf_{uij} = r^{\a+2} C_{ij} \, , \quad
\vf_{ijk} = r^{\a+3} B_{ijk} \, , 
\ee
fits nicely in the form of the equations $\cF_{\m\n\r} = 0$. For instance, it allows to cancel among each other the contributions from the two terms entering the equations imposed by the radial components of the Fronsdal tensor. Substituting \eqref{ansatz-alpha} in \eqref{R_ruu}--\eqref{R_rij} one indeed obtains
\begin{align}
\cF_{ruu} & = r^{\a-2} \left\{ \a(\a+n-1)\, B - (\a+1)\, D^i U_i \right\} + \cO(r^{\a-3}) \, , \\[5pt]
\cF_{rui} & = r^{\a-1} \left\{ (\a-1)(\a+n)\, U_i - \a\, D^j C_{ij} \right\} + \cO(r^{\a-2}) \, , \\[2pt]
\cF_{rij} & = r^{\a} \left\{ (\a-2)(\a+n+1)\, C_{ij} - (\a-1)\, D^k B_{ijk} \right\} + \cO(r^{\a-1}) \, .
\end{align}
Eq.~\eqref{R_ijk} gives instead
\be
\cF_{ijk} = -\, r^{\a+2}\, (2\a+n)\, \pr_u B_{ijk} + \cO(r^{\a+1}) \, ,
\ee
so that the corresponding equation of motion is satisfied at leading order if \mbox{$\a = - n/2$} (which avoids the restrictive condition $\pr_u B_{ijk} = 0$ having no analogue in gravity). This leads to the fall-off conditions considered in Section~\ref{sec: rot}. When $n > 2$ the equations of motion associated to \eqref{R_ruu}--\eqref{R_rij} are then satisfied at leading order provided that 
\be \label{rel-tens-2}
B = \frac{2}{n}\, D^i U_i \, , \qquad
U_i = \frac{2}{n+2}\, D^j C_{ij} \, , \qquad
C_{ij} = \frac{2}{n+4}\, D^k B_{ijk} \, .
\ee
Imposing these conditions, the equations of motion associated to \eqref{R_uuu}--\eqref{R_uij} are satisfied as well at leading order, consistently with the number of constraints imposed by the Bianchi identities (which is equal to the number of independent components in $\cF_{rrr}$, $\cF_{rru}$, $\cF_{rri}$, $\cF_{uuu}$, $\cF_{uui}$ and in the identically traceless $\cF_{uij}$).
The case $n = 2$ is special also in this respect, since the equation $\cF_{ruu} = \cO(r^{\a-3})$ is identically satisfied for $\a = - 1$ thanks to the factorisation of $(\a+1)$. As discussed in Section~\ref{sec: Bondi}, in this case one should look at 
\be
\cF_{uuu} = 3\,r^{-2} \left\{ \pr_u B - \pr_u D^i U_i \right\} + \cO(r^{-3}) \, ,
\ee
which gives the weaker relation $\pr_u B = \pr_u D^i U_i$ instead of the first condition in \eqref{rel-tens-2}.

\section{More details on the residual gauge symmetry}\label{sec:details}

In this appendix we display the differential combinations of the tensors $T$, $\r^i$ and $K^{ij}$ that we omitted in Section~\ref{sec: rot} when presenting the structure of the gauge parameters generating asymptotic symmetries. We also display the identities that allow to fully express the variations \eqref{phi_uij} and \eqref{phi_ijk} in terms of the differential constraints $\cK_{ijk}$, $\cR_{ijk}$ and $\cT_{ijk}$ when $n>2$. Finally, we show that the number of solutions of \eqref{killing} coincides with that of \eqref{constr-KR} and \eqref{constr-T}, at least when solving the latter in their flat limit. 

\subsection*{Gauge parameters}

The tensors entering the expansion of $\e^{ij}$ in \eqref{e^ij} are
\begin{align}
\cT_2^{ij} & = - \frac{1}{n+2} \left( D^{(i} \Ddot K^{j)} - \frac{2}{n+1}\, \g^{ij} \Ddot\Ddot K \right) , \\
\cT_4^{ij} & = \frac{1}{4(n+1)(n+2)} \left( D^{(i} D^{j)} + 4\, \g^{ij} \right) \Ddot\Ddot K \, , \\
\cU_1^{ij} & = D^{(i} \r^{j)} - \frac{2}{n+1}\, \g^{ij} \Ddot \r \, , \\
\cU_3^{ij} & = - \frac{1}{2(n+1)} \left( D^{(i} D^{j)} - \frac{2}{n+2}\, \g^{ij}\! \left( \Box - 2 \right) \right) \Ddot \r \, , \\
\cV_2^{ij} & = - \frac{1}{4}  \left( D^{(i} D^{j)} - \frac{2}{n+2}\, \g^{ij}\! \left( \Box - 2 \right) \right) T \, .
\end{align}
The tensors entering the expansion of $\e^{ri}$ in \eqref{e^ri} are
\begin{align}
\cA^{i} & = \Ddot K^i + \frac{u}{r} \left( \Ddot K^i  - \frac{1}{2(n+1)} D^i \Ddot\Ddot K \right) - \frac{(u\,r^{-1})^2}{2(n+1)}\, D^i \Ddot\Ddot K \, , \\
\cB^{i} & = \left( \Box + (n-1) \right) \r^i + \frac{n}{n+1}\, D^i \Ddot \r - \frac{u\,r^{-1}}{n+1}\, D^i\! \left( \Box + n \right) \Ddot \r \, .
\end{align}

\subsection*{Gauge variations}

The variations $\d \vf_{uij}$ and $\d \vf_{ijk}$ displayed, respectively, in \eqref{phi_uij} and \eqref{phi_ijk} contain some terms that have not been rewritten in terms of the differential constraints \eqref{constr-KR} and \eqref{constr-T}. This reflects the existence of non-trivial variations of the boundary data in $n=2$ under both higher-spin supertranslations and superrotations.\footnote{Variations of the boundary data in the linearised theory signal the presence of central charges in the classical algebra of asymptotic symmetries. See {\it e.g.}\ Section~2.4 of \cite{charges-AdS} for a discussion in the AdS$_3$ case.} On the other hand, when $n>2$, compatibility with the falloffs \eqref{boundary} requires that these residual variations vanish. This is guaranteed by the following identities:
{\allowdisplaybreaks
\begin{align}
& \frac{n(n-2)}{2(n+1)(n+2)} \left( D_{(i} D_j D_{k)} + 8\, \g_{(ij} D_{k)} \right) \Ddot\Ddot K \nn \\*
& = \Box^2 \cK_{ijk} - \Box D_{(i} \Ddot \cK_{jk)} + \frac{n}{2(n+1)}\, D_{(i} D_j \Ddot\Ddot \cK_{k)} + 3\, D_{(i} \Ddot \cK_{jk)} + (n-6) \Box \cK_{ijk} \nn \\*
& - \frac{1}{n+1}\, \g_{(ij|} \left( D_{|k)} \Ddot\Ddot\Ddot \cK - \Box \Ddot\Ddot\cK_{|k)} + (n+3) \Ddot\Ddot\cK_{|k)} \right) -3(n-3)\, \cK_{ijk} \, , \label{identity_1}\\[10pt] 
& \frac{n-2}{2} \left( D_{(i} D_j D_{k)} - \frac{2}{n+2}\, \g_{(ij} D_{k)}\! \left( 3\Box + 2 (n-1) \right) \right) \Ddot\r \nn \\*
& =  -\,(n+1)\, \Box \cR_{ijk} + \frac{n+2}{2}\, D_{(i}\Ddot\cR_{jk)} - \g_{(ij} \Ddot\Ddot \cR_{k)} + 3(n+1)\, \cR_{ijk} \, , \label{identity_2} \\[10pt]
& \frac{3(n-2)}{n+2} \left( D_{(i} D_{j)} + 2\,\g_{ij} \right) \left( \Box + 2(n+1) \right) \Ddot \r \nn \\*
& = \Box \Ddot \cR_{ij} - D_{(i} \Ddot\Ddot \cR_{j)} + \frac{1}{n+1}\, \g_{ij} \Ddot\Ddot\Ddot \cR - 2\, \Ddot\cR_{ij} \, .\label{identity_3}
\end{align}
}
Note the overall factors $(n-2)$ in the left-hand sides, which still allow for non-trivial variations in four space-time dimensions.

\subsection*{Differential constraints}

Let us now turn to the equations \eqref{constr-KR} and \eqref{constr-T}. For simplicity, we evaluate their number of independent solutions by analysing them in the flat limit, assuming that the dimension of the solution space remains the same, as it is manifest for the equation $\cK^{ijk} = 0$, which is only rescaled under Weyl rescalings of the metric. We therefore consider the equations 
\begin{align}
\pr^{(i} K^{jk)} - \frac{2}{n+2}\, \h^{(ij} \pr\cdot K^{k)} & = 0 \, , \label{conformal-flat-1} \\[5pt]
\pr^{(i} \pr^j \r^{k)} - \frac{1}{n+2}\, \h^{(ij|}\! \left(\, \Box \r^{|k)} + 2\,\pr^{|k)} \prd\r \,\right) & = 0 \, , \label{conformal-flat-2} \\[5pt]
\pr^{i} \pr^j \pr^{k} T - \frac{1}{n+2}\, \h^{(ij} \pr^{k)} \Box T & = 0 \, . \label{conformal-flat-3}
\end{align}
The different relative factors with respect to \eqref{K}--\eqref{T} are induced by our convention for the symmetrisations, since $\pr^{(i} \pr^j \r^{k)}$ contains {\it e.g.}\ less terms than $D^{(i} D^j \r^{k)}$ because ordinary derivatives commute. When $n > 2$, eqs.~\eqref{conformal-flat-1}--\eqref{conformal-flat-3} are solved by
{\allowdisplaybreaks
\begin{align}
K_{ij} & = a_{ij} + \left( b_{(i}x_{j)} - \frac{1}{n}\, \h_{ij} b\cdot x\right) + \o_{ij|k}\,x^k + \l \left( x_i x_j - \frac{1}{n}\, \h_{ij} x^2 \right) + \r_{k|(i}x_{j)}x^k + \O_{ij|kl} x^k x^l \nn \\* 
& + \left(  2\, c_{k(i} x_{i)}x^k  - c_{ij}\, x^2 - \frac{2}{n}\, \h_{ij} c_{kl} x^k x^l \right) + \left( 2\,\tilde{b}_k x_ix_jx^k - \tilde{b}_{(i} x_{j)}x^2 - \frac{1}{n}\, \h_{ij} (\tilde{b}\cdot x) x^2 \right) \nn \\*
& + \left( 2\,\tilde{\o}_{kl|(i}x_{j)}x^kx^l \! + \tilde{\o}_{ij|k}x^kx^2 \right) + \tilde{c}_{kl} \left( 4\,x_ix_jx^kx^l - 4\, \n^k{}_{(i} x_{j)} x^l x^2 + \d_{i}{}^k \d_{j}{}^{l} x^4  \right) , \label{K-sol-gen} \\[5pt]
\r^i & = a_i + \l\,x_i + b_{ij}x^j + \o_{i|j}x^j + \tilde{a}_i x^2 + c_jx_ix^j + \O_{jk|i}x^jx^k + \tilde{\l}\, x_i x^2 + \tilde{\o}_{i|j}x^j x^2 \nn \\*
& + \tilde{b}_{jk} \left( x_ix^jx^k - \frac{1}{2}\, \d_i{}^j x^kx^2\right) + \tilde{c}_j \left( x_i x^j x^2 - \frac{1}{2}\, \d_i{}^j x^4 \right) , \label{r-sol-gen} \\[5pt]
T & = a + b_i x^i + c_{ij} x^ix^j + \l\, x^2 + \tilde{b}_i x^i x^2 + \tilde{\l}\, x^4 \, , \label{T-sol-gen}
\end{align}
}
\!where all tensors in the solutions are traceless and irreducible. As a result, the number of integration constants is
\be
T \sim \frac{(n+1)(n+4)}{2} \, , \
\r^i \sim \frac{n(n+2)(n+4)}{3}\, , \
K^{ij} \sim \frac{(n-1)(n+2)(n+3)(n+4)}{12}\, , 
\ee 
corresponding to the number of independent components, respectively, of a $\{2\}$, a $\{2,1\}$ and a $\{2,2\}$ Young-projected and traceless tensors in $n+2$ dimensions, where each number in the list denotes the length of a row in the Young tableau. The sum matches the number of integration constants in \eqref{sol-killing}, which is related to the components of a $\{2,2\}$-projected traceless tensor in $n+3$ dimensions. When $n=2$, locally eqs.~\eqref{conformal-flat-1} and \eqref{conformal-flat-2} are instead solved by \eqref{sol2-K} and \eqref{sol2-r}, while globally well-defined solutions can still be cast in the form \eqref{K-sol-gen} and \eqref{r-sol-gen}.

\end{appendix}



\end{fmffile}
\end{document}